\documentclass[12pt]{iopart}
\usepackage{iopams}  
\usepackage{braket}
\usepackage{amsthm}
\usepackage{float}
\usepackage{hyperref}
\usepackage{graphicx}
\usepackage{subfig}

\usepackage{latexsym}
\usepackage{bbold}
\usepackage{bbm}
\usepackage{url}

\newcommand{\prob}{{\bf Pr}}
\newtheorem{thm}{Theorem}[section]

\newtheorem{lem}[thm]{Lemma}
\newtheorem{prop}[thm]{Proposition}

\theoremstyle{remark}

\theoremstyle{definition}
\newtheorem{defi}[thm]{Definition}

\begin{document}

\title{Space-Time Circuit-to-Hamiltonian Construction and Its Applications}
\author{Nikolas P. Breuckmann and Barbara M. Terhal}
\address{Institute for Quantum Information, RWTH Aachen University, 52056 Aachen, Germany}\eads{\mailto{breuckmann@physik.rwth-aachen.de},\mailto{terhal@physik.rwth-aachen.de}}

\date{\today}
\begin{abstract}
The circuit-to-Hamiltonian construction translates dynamics (a quantum circuit and its output) into statics (the groundstate of a circuit Hamiltonian) by explicitly defining a quantum register for a clock.  The standard Feynman-Kitaev construction uses one global clock for all qubits while we consider a different construction in which a clock is assigned to each interacting qubit. This makes it possible to capture the spatio-temporal structure of the original quantum circuit into features of the circuit Hamiltonian. The construction is inspired by the original two-dimensional interacting fermion model in \cite{MMC:groundstate}. We prove that for one-dimensional quantum circuits the gap of the circuit Hamiltonian is appropriately lower-bounded so that the applications of this construction for QMA (and partially for quantum adiabatic computation) go through. For one-dimensional quantum circuits, the dynamics generated by the circuit Hamiltonian corresponds to diffusion of a string around the torus. 
\end{abstract}

%\maketitle

%\makeatletter
%\@starttoc{toc}
%\makeatother

% \tableofcontents

%%%%%%%%%%%%%%%%%%%%%%%%%%%%
% Introduction
%%%%%%%%%%%%%%%%%%%%%

\section{Introduction}
\label{sec:intro}

In \cite{feynman:qmc} Feynman considered how to simulate a quantum circuit by unitary dynamics generated by a time-independent Hamiltonian $H$. Imagine that the quantum circuit consists of $L$ unitary gates $U_1, \ldots, U_L$ on $n$ qubits. Feynman's idea was to introduce a clock-register $\ket{t}$ with time $t$ running from $t=0$ to $L$ such that for each unitary gate $U_{t}$ in the circuit, we have a term $H_t$ in the Hamiltonian $H$, i.e.
\begin{equation*}
H_t=U_t \otimes \ket{t}\bra{t-1}+U_t^{\dagger} \otimes \ket{t-1}\bra{t},\;\; H=\sum_{t=1}^{L} H_t.
\end{equation*}
Alternatively, one can construct a Hamiltonian $H_{circuit}$ such that the groundstate of $H_{circuit}=\sum_{t=1}^L H_t$ is the {\em history state} of the quantum circuit \cite{KSV:computation}. We then take \footnote{Sometimes a prefactor of $\frac{1}{2}$ is included to make $H_t$ a projector.}
\begin{equation*}
H_t=-U_t \otimes \ket{t}\bra{t-1}-U_t^{\dagger} \otimes \ket{t-1}\bra{t}+\ket{t}\bra{t}+\ket{t-1}\bra{t-1} \geq 0.
\nonumber
\end{equation*}
The zero energy groundstate of the {\em circuit Hamiltonian} $H_{circuit}$ is 
\begin{equation*}
\ket{\psi_{history}}=\frac{1}{\sqrt{L+1}}\sum_{t=0}^L U_t \ldots U_1 \ket{\xi} \otimes \ket{t},
\end{equation*}
for any input state $\ket{\xi}$ to the circuit. It is not hard to analyze the spectrum of $H_{circuit}$ as one can transform the dependence on the specific gates $U_1, \ldots, U_L$ away by a unitary transformation $W=\sum_{t=0}^{L} U_t \ldots U_1 \otimes \ket{t}\bra{t}$ such that $W^{\dagger} H_{circuit}(U_1,\ldots, U_L) W=H_{circuit}(U_1=I,\ldots,U_L=I)$. 
This unitarily-transformed circuit Hamiltonian corresponds to that of a particle (whose location is $t$) moving on a 1D line:  the eigenvalues of $H_{circuit}$ are $\lambda_k=2(1- \cos q_k)$ with $q_k=\frac{\pi k}{L+1}$ for $k=0,\ldots, L$. The gap above the ground-space of $H_{circuit}$ is thus easily lowerbounded as $\Omega(L^{-2})$, corresponding to the lowest $k \neq 0$ eigenstate.
If one is given the history state, one can measure the clock register $t$ and, with probability $1/(L+1)$, obtain the output of the quantum circuit. In order to increase the probability of getting the output to some constant, one can pad the quantum circuit with, say, $L$ identity gates at the end, so that the probability of measuring any time $t \in [L,2L]$ is approximately $1/2$. For all times in this interval, the qubits are in the output state of the quantum circuit. It has been shown how the circuit-to-Hamiltonian construction can be used directly as a model for universal quantum adiabatic computation \cite{ADLLKR:adia}. 

The circuit-to-Hamiltonian construction was first used by Kitaev in quantum complexity theory to prove that certain problems are QMA-complete. The complexity class QMA (Quantum Merlin Arthur) \cite{KSV:computation} is the quantum equivalent of the class NP (or its probabilistic variant MA). Informally, in QMA the classical proof or witness and the classical verifier of NP are replaced by a quantum proof $\ket{\xi}$ and a quantum verifier. The formal definition is

\begin{defi}[QMA \cite{KSV:computation,watrous:qma}]
A promise problem $L=L_{yes}\cup L_{no}\subseteq \{0,1\}^*$ belongs
to QMA iff there exist a polynomial $p(n)$ and a polynomial-time generated family of quantum circuits $\{C_n\}$ which take an input of $n+p(n)$ qubits such that such that for all $n$ and all $x \in \{0,1\}^n$, 
\begin{eqnarray*}
x\in L_{yes} &\Rightarrow & \exists \, \xi, \quad
\prob{\left[C_n(x,\xi)=1\right]}\ge 2/3,\;\;  \mbox{\rm (Completeness)} \label{comp} \\
x\in L_{no} &\Rightarrow & \forall\, \xi, \quad
\prob{\left[C_n(x,\xi)=1\right]}\le 1/3. \;\; \mbox{\rm (Soundness)} 
\label{sound}
 \end{eqnarray*}
 where $\xi$ is a $p(n)$-qubit quantum state.
\label{def:QMA}
\end{defi}

The completeness and soundness errors $(\frac{2}{3},\frac{1}{3})$ can be amplified to $(1-\epsilon, \epsilon)$ where $\epsilon=2^{-{\rm poly}(n)}$ \cite{KSV:computation, MW:error}, thus making these errors exponentially small, without increasing the number of qubits of the witness $\xi$.

To prove that a computational (promise) problem is QMA-complete, one needs to prove that (1) the problem is contained in the complexity class QMA and (2) that the problem is QMA-hard.  The general `local Hamiltonian' problem has been shown to be in QMA, e.g. 

\begin{prop}[\cite{KSV:computation}]
Let $H=\sum_i H_i$ be a Hamiltonian on $n$ qubits with $||H_i||=O(1)$ and each $H_i$ acts on $O(1)$ qubits non-trivially.
We have the following promise: either there exists a state $\psi$, $\bra{\psi} H \ket{\psi} \leq a$ ({\rm YES}) or $\forall \psi, \bra{\psi} H \ket{\psi} \geq b$ ({\rm NO}) for some given $a,b$ (described by some ${\rm poly}(n)$ bits) with $|a-b| \geq \frac{1}{{\rm poly}(n)}$. The problem of deciding between YES and NO is in the class {\rm QMA}.
\end{prop}

The idea behind the containment in QMA is simple: if YES, Merlin (the prover) can give Arthur (the verifier) a ground-state and Arthur can estimate the energy of this state with $1/{\rm poly}(n)$ precision using an efficient quantum circruit. If this answer is NO, then Merlin cannot give any state which has low enough energy to fool Arthur.

Using the circuit-to-Hamiltonian construction, Kitaev proved that $5$-local Hamiltonian problem (where each $H_i$ acts on at most $5$ qubits) is QMA-complete \cite{KSV:computation}. Since then, many variants of the local Hamiltonian problem have been shown to be QMA-complete such as 1D local Hamiltonians \cite{AGIK:1d}. See \cite{GN:qma1, CM:qma} and references therein for the most recent results. Various new results for QMA-complete problems have so far come about by modifications of the circuit-to-Hamiltonian construction, different realizations of clocks and the use of perturbation gadgets \cite{OT:qma}.

In this paper we will show how a different circuit-to-Hamiltonian construction, the space-time circuit-to-Hamiltonian construction (see \cite{Margolus} for early work on this construction), can be used to give QMA-completeness results. In the next section we review a modification of the Feynman-Kitaev construction with circular time. In Section \ref{sec:stcircuit} we will present the space-time circuit-to-Hamiltonian construction for general quantum circuits.  In Section \ref{sec:relateMLM} we show how the space-time circuit-to-Hamiltonian construction for one-dimensional quantum circuits relates to a two-dimensional fermionic model which has been previously proposed as a model for adiabatic computation. In Section \ref{sec:STcirctime} we show how to modify the space-time construction for circular time: this is convenient for our later mathematical analysis. In Section \ref{sec:gap} we start with a spectral analysis of the circuit Hamiltonian and we focus our attention on one-dimensional quantum circuits between nearest neighbor qubits in Section \ref{sec:FMHeis}. An important result in Section \ref{sec:FMHeis} is the mapping of the Hamiltonian dynamics onto that of a diffusing string. The string can be parametrized by internal variables determining the shape of the string (dynamics of a Heisenberg model) and an arbitary boundary point which is moving on a one-dimensional line. This mapping allows us to lower bound the spectral gap of the circuit Hamiltonian. The results in this Section \ref{sec:FMHeis} then play an important role in Section \ref{sec:QMA} where we prove, loosely speaking, that determining the ground-state energy of a 2D interacting fermion model with a specific constraint on the fermion number is QMA-complete. In Section \ref{sec:fermrevis} we consider the consequence of our results for quantum adiabatic computation.

We present the space-time circuit-to-Hamiltonian construction in its generality as we believe that the association of a Hamiltonian with a quantum circuit may in the future have other applications beyond the one directly discussed here. 

\subsection{Circular Time}

\begin{figure}[htb]
\centering
\includegraphics[width=0.5\hsize]{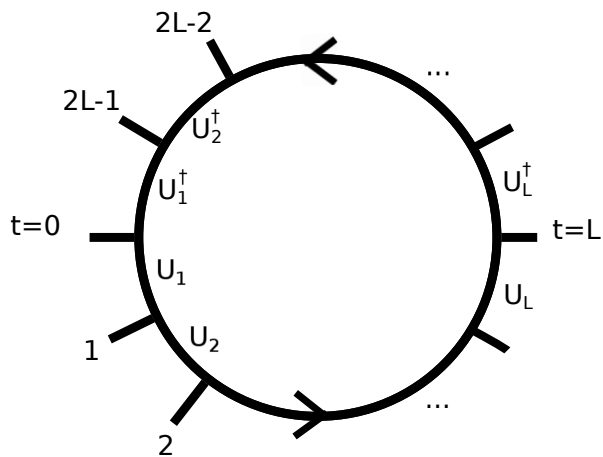} 
\caption{Representation of the Feynman-Kitaev circuit-to-Hamiltonian construction with circular time \cite{nagaj:circular}. At $t=L$, the qubits are in the output state of the quantum circuit while evolving further along the circle will undo the evolution. The evolution from any point, say $t=0$, to another point $t$ on the circle is well-defined, even though the evolution can happen via two different paths.}
\label{fig1}
\end{figure}

For any quantum circuit one can define a circuit Hamiltonian whose dynamics correspond to a particle moving on a {\em circle} instead of a line (see \cite{nagaj:circular}). We will use this idea in this paper as it is easier to analyze, so let us give some details, see Fig.~(\ref{fig1}). We define a {\em circular} clock register $t=0,\ldots 2L-1$ where we identify $t=2L$ with $t=0$ ($t \in Z_{2L}$). The idea is to use the sequence of unitary gates $U_1, \ldots, U_L$ of the quantum circuit for the two different ways one can go from $t=0$ to the opposite point on the circle, $t=L$, see Fig.~(\ref{fig1}). More generally, we define some new, yet to be specified, gates $U_{L+1}, \ldots U_{2L}$ and take as before
\begin{equation*}
t \in [1,2L] \colon \; H_t=-\left(U_t \otimes \ket{t}\bra{t-1}+h.c.\right)+\ket{t}\bra{t}+\ket{t-1}\bra{t-1}. 
\label{eq:circ_time}
\end{equation*}
Let $H_{circuit}=\sum_{t=1}^{2L} H_t$. As $H_{circuit}$ is a sum of positive-semidefinite operators, it only has a zero energy if all terms $H_t$ have zero energy. W.l.o.g. we can take the groundstate to be of the form $\sum_{t=0}^{2L-1} \ket{\psi_t} \ket{t}$ which is a zero energy state if and only if
\begin{equation*}
t \in [1,2L]\colon\; \ket{\psi_{t}}=U_t \ket{\psi_{t-1}}.
\end{equation*}
This implies that the unitary evolution from a state $\ket{\psi_t}$ around the entire circle must act as $I$ on the state $\ket{\psi_t}$. Equivalently, we have $U_{2L} \ldots U_{L+1} U_L \ldots U_1 \ket{\xi}=\ket{\xi}$
where $\ket{\xi}=\ket{\psi_{t=0}}$. Depending on the choice for $U_{L+1}, \ldots, U_{2L}$, this defines a {\em subspace} of states $\ket{\xi}$. When we choose $U_t=U_{2L-t+1}^{\dagger}$ for $t=L+1,\ldots, 2L$, the subspace $\ket{\xi}$ is the whole space and the history state of the circuit is 
\begin{equation}
\ket{\psi_{history}}=\frac{1}{\sqrt{2L}} \sum_{t=0}^{2L-1} U_t \ldots U_2 U_1 \ket{\xi} \otimes \ket{t}, \forall\;\xi
\label{his:circ}
\end{equation}
where the latter part (for $t > L$) of the evolution unravels the earlier part.  An additional observation is that if the original quantum circuit contains some $I$ gates here and there, then the gates need not explicitly be included in the unraveling evolution, in order for there to be a zero energy history state for any $\xi$.  

%Thus for such choice of $\{U_i\}_{i=L}^{2L}$, $H_{circuit}$ describes a {\em unique} evolution from any time $t$ to any %other time $s$, while for other (arbitrary) choices of $U$ the paths going left or right around the circle may correspond to %different unitary evolutions. 

% One can imagine that this circular construction is also useful in constructing Hamiltonians for circuits of which the input to % the circuit is, for some reason, constrained, for example it is a classical bitstring as in the definition of the complexity class % QCMA.

Note that the history state of this circular time construction, Eq.~(\ref{his:circ}), contains the output of the original circuit when we measure time and find $t=L$. As before, we can pad the original circuit with $I$ gates at the end such that we have a window of time around $t=L$ when the qubits are in the output state of the original quantum circuit. Hence, if one is given (a fast adiabatic path towards) the ground-state of the circuit Hamiltonian, one can measure the output of the quantum circuit with such circular-time model similar as in the linear-time model.

%for example, classical bit string. We could then take $\tilde{U}_{L+p} =Z U_{L-p}^{\dagger}$ for those gates $U_{L-p}$ %in the original circuit which are the {\em first} gates to act on the input qubits. 
%This choice imposes the condition that the state $\ket{\psi_{t=0}}=\ket{\xi}$ is invariant under Pauli $Z$ on all the qubts

%NB
%Before we introduce the novel circuit-to-Hamiltonian construction, we remark that the circuit-to-Hamiltonian construction has an intimate connection with the cosmology ideas of Page and Wootters \cite{PW:cosmo, moreva+:TE} which, in our language, posit that there is no observable difference between a universe which evolves according to the circuit dynamics versus a universe which simply finds itself in the groundstate of the circuit Hamiltonian $H_{circuit}$. 

\subsection{Space-Time Circuit-to-Hamiltonian Construction}
\label{sec:stcircuit}

We consider a quantum circuit on $n$ qubits with single and two-qubit gates $U_{i},i=1,\ldots, S$ where $S$ is the {\em size} of the circuit. As some gates can be executed in parallel on different qubits, the circuit also has a certain {\em depth} $D \leq S$. The circuit may have a geometric structure, i.e. only nearest-neighbor qubits on some $d$-dimensional lattice or space interact. The space-time circuit-to-Hamiltonian defines a circuit Hamiltonian $H_{circuit}$ whose properties relate to the geometric structure and the depth $D$ of this quantum circuit. 

%NB changed wrong subscript U_h
Each gate $U_i$ in this circuit will correspond to a term in $H_{circuit}$. The gates can be labeled as $U_t^1[q]$ for a single-qubit gate acting at time-step (depth) $t=1,\ldots,D$ on qubit $q$, or a two-qubit gate $U_t^2[q,p]$ acting at time-step $t$ on qubits $q$ and $p$.

The construction that we will analyze later has circular time, see Sec.~\ref{sec:STcirctime}, but for simplicity we first define the model with linear time.  For {\em each} qubit $q$ in the original circuit, we define a clock register $\ket{t}_q$ with $t=0, \ldots, D$. Thus the global clock in the Feynman-Kitaev construction gets replaced by a {\em time-configuration} $\ket{t_1,\ldots,t_n}_{1,\ldots n}$. Consider a single qubit gate $U^1_t[q]$ acting on qubit $q$ at time-step $t$ in the quantum circuit. For each such gate, there is a term $H_t^1[q]$ in $H_{circuit}$ of standard form, i.e. 
\begin{equation*}
H^1_t[q]=-\left( U^1_t[q] \otimes \ket{t}\bra{t-1}_q +h.c.\right)+\ket{t}\bra{t}_q+\ket{t-1}\bra{t-1}_q.
\end{equation*}
Clearly, if the quantum circuit were to consist of single qubit gates only, the history state would be a tensor product of history states, one for each qubit independently. In such a scenario, the clocks of the qubits can be completely unsychronized and measure different times.

For every two qubit gate $U^2_t[q,p]$ acting on qubits $p$ and $q$ at time $t_q=t_p=t$ in the quantum circuit, we have  in $H_{circuit}$ the term 
\begin{eqnarray}
H^2_t[q,p]=-\left(U^2_t[q,p] \otimes \ket{t,t}\bra{t-1,t-1}_{q,p}+h.c.\right)
\nonumber \\+\ket{t,t}\bra{t,t}_{q,p}+\ket{t-1,t-1}\bra{t-1,t-1}_{q,p} \geq 0.
\label{eq:2q}
\end{eqnarray}
Note that $H^2_t[q,p]$ always has zero energy when the clocks of qubits $q$ and $p$ measure unequal times. We take $H_{circuit}=\sum_{t=1}^D H_t$ where $H_t$ is a sum over all $H^2_t[q,p]$ and $H^1_t[q]$ for various $q,p$, corresponding to gates $U^2_t[q,p]$ and $U^1_t[q]$ which act in parallel at time $t$.

\subsection{Valid Time-Configurations}
We consider the zero energy states of this circuit Hamiltonian. First we define what we call {\em invalid} time-configurations $\ket{t_1,\ldots,t_n}$. Invalid configurations are the time-configurations in which, of at least one pair of qubits, say, the pair $(q,p)$ which interacts in some two-qubit gate $U^2_t[q,p]$ in the quantum circuit, it holds that either $(t_q < t) \wedge (t_p \geq t)$ or $(t_p < t) \wedge (t_q \geq t)$. Informally, this means that one qubit has gone through the gate while its partner qubit has not yet gone through the gate. If one would evolve with $H_{circuit}$ starting from the all-synchronized state $\ket{t_1=0,\ldots,t_n=0}\otimes \ket{\xi}$, then clearly the resulting state would not have any support on invalid time-configurations as qubits always go together through two-qubit gates by Eq.~(\ref{eq:2q}). Stated differently, $H_{circuit}$ preserves the space of valid time-configurations and its eigenstates split into a sectors of valid and invalid eigenstates. 

On the space of invalid time-configurations, one can easily find zero energy eigenstates for $H_{circuit}$, but we will not be interested in these states. If we apply this construction for quantum adiabatic computation, Section \ref{sec:fermrevis}, we can start our adiabatic computation in the space of valid time-configurations and thus remain in this subspace. If we apply the construction to QMA, we need to do some additional work, see Section \ref{sec:QMA}.

We consider zero energy states in the space of valid time-configurations. We restrict ourselves to quantum circuits which only employ two-qubit gates \footnote{Single-qubit gates can always be absorbed into two-qubit gates. The presence of single-qubit gates would lead to some differences, for example the presence of gapped excitations in $H_{circuit}$ which are localized in space-time.}.
For such quantum circuits, a valid time-configuration $\ket{t_1,\ldots,t_n}$ has zero energy when, for {\em every} two-qubit gate $U^2_t[q,p]$ in the circuit, the clock-times $t_q$ and $t_p$ are either $t_q \neq t_p$, or $t_p=t_q \notin\lbrace t-1,t\rbrace$ as then each term $H^2_t[q,p]$ has zero energy with respect to $\ket{t_1,\ldots, t_n}$. Such configurations do not evolve and we could call these configurations {\em light-like}. More precisely, assume we give each qubit $q$ a spatial location $r_q$, all points being equidistant. Then the valid time-configurations $(t_1,t_2,\ldots, t_n)$ are such that each pair $(r_q,t_q)$ and $(r_p,t_p)$ of space-time points of this configuration are either space-like separated or light-like separated, as there is no causal relation between such pairs of points $(r_q,t_q)$ and $(r_p,t_p)$ in the original circuit. The invalid configurations are such that at least one pair of points of this configuration is time-like separated. One cannot associate a metric with such discrete circuit directly, but in the continuum limit the causal cones of qubits in the quantum circuit gives rise to a (uniform) 2D Minkowski metric.

\begin{figure}[ht]

   \centering
      \subfloat[]{\includegraphics[scale=0.2]{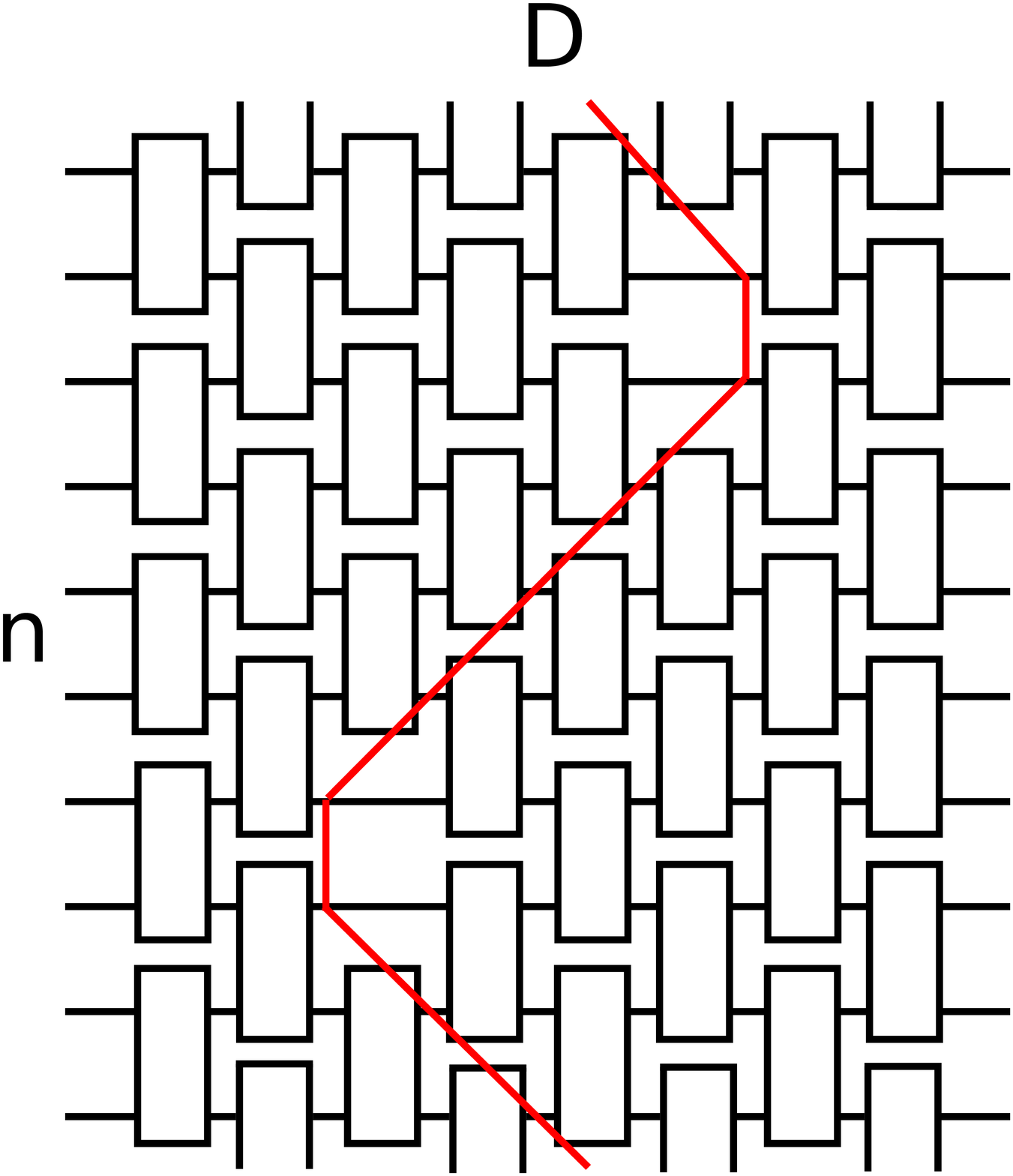}}\qquad%NPB space between subfloats
      \subfloat[]{\includegraphics[scale=0.2]{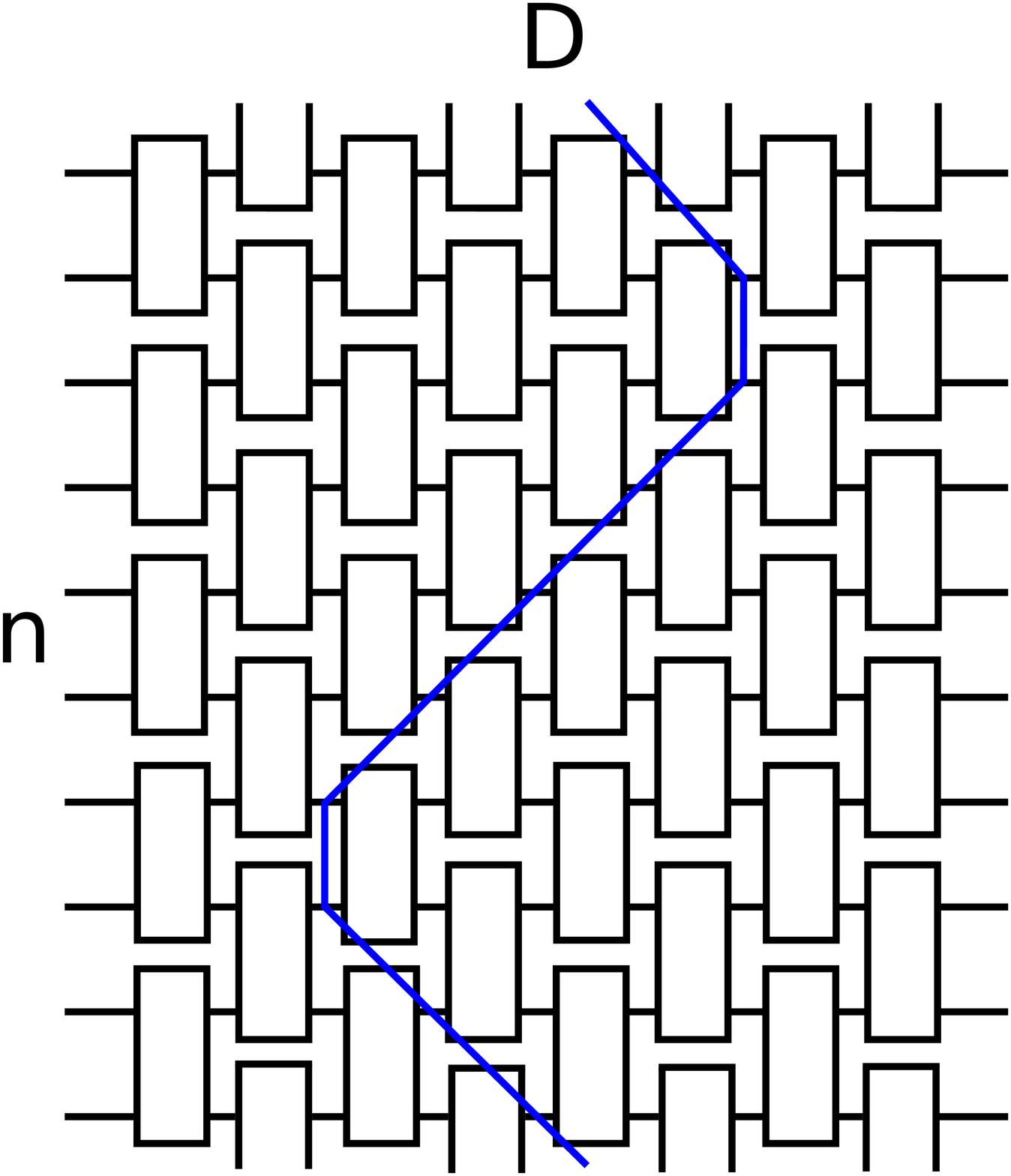}}
       \caption{(a) One-dimensional quantum circuit on $n$ qubits and depth $D$ where the (red) line indicates a zero energy time configuration. (b) One-dimensional quantum circuit on $n$ qubits with nearest-neighbor interactions on a circle and depth $D$ ($n$ and $D$ both even) which is analyzed in this paper. The (blue) line is not a zero energy configuration but evolves under $H_{circuit}$.}
\label{fig2}
\end{figure}

Let us illustrate these notions with quantum circuits that will mostly concern us, namely one-dimensional quantum circuits with nearest-neighbor qubits interacting in two-qubit gates, depicted in Fig.(\ref{fig2}). The quantum circuit in Fig.(\ref{fig2})(a) has a beginning and an end and periodic boundary condition in space, but  some two-qubit gates are missing in the circuit so that the (red) line represents a zero energy configuration. The quantum circuit in Fig.~(\ref{fig2})(b) has no zero energy configurations. Note that $n$ and $D$ are both even. Fig.~(\ref{fig3}) is an example of a quantum circuit with periodic boundary conditions in both space and time which does have unavoidable zero energy configurations, see Section \ref{sec:STcirctime}.

%NB changed ll to zero energy
For quantum adiabatic computation, the valid zero energy configurations are harmless as we can avoid starting the computation in such non-evolving configurations. For the application to QMA, the existence of valid zero energy configurations must be avoided as the goal is to construct a Hamiltonian where the existence of a zero energy groundstate depends on the computation done by the quantum circuit. If there are valid zero energy configurations, it is not clear how to modify $H_{circuit}$ to make such configurations have non-zero energy. As we see, it is simple to avoid zero energy configurations by ensuring that the quantum circuit has two-qubit and single-qubit (possibly $I$) gates throughout which propagate the clocks.

%Let us briefly comment on how the presence of single qubit gates modifies the notion of light-like configurations. If single qubit gates are interspersed in between two qubit gates, then zero energy configurations are no longer simple time-configurations. For each interval $t_q \in [t_a,t_b]$ with only single qubit gates on a particular qubit $q$, let $\ket{\psi_h^{t_a,t_b}[q]}$ be the local history state of that qubit in that interval. If those single-qubit gates were absent, then let the $(t_b-t_a)$ configurations $t_1,\ldots t_{q-1},t_q,t_{q+1},\ldots,t_n$ with with $t_q=t_a,t_a+1, \ldots, t_b$ be light-like. Then with those gates present, the single corresponding configuration $\ket{t_1,t_{q-1},\psi_h^{t_a,t_b}[q], t_{q+1},\ldots, t_n}$ will be light-like.

\subsection{Relation with the fermionic ground-state model of \cite{MMC:groundstate, MMC:scaling, MLM:qadiabatic}}
\label{sec:relateMLM}

In \cite{MMC:groundstate} the authors formulate a (fermionic) model which allows for universal quantum computation by adiabatically modifying a circuit Hamiltonian \cite{MLM:qadiabatic}. Imagine we have a quantum circuit on $n$ qubits, e.g. the one in Fig.~(\ref{fig2})(b), of depth $D$. With every qubit $q$, we associate $2(D+1)$ fermionic modes with creation operators $a^{\dagger}_t[q], b^{\dagger}_{t}[q], t=0,\ldots, D$. One can view these $2n (D+1)$ modes as the state-space of $n$ spin-$1/2$ fermions, where each fermion can be localized at sites on a one-dimensional (time)-line of length $D+1$. The spin-state of the $n$ fermions represents the state of the computation while the clock of each qubit is represented by where the fermion is on the one-dimensional line. Let  $C_{t}[q]=\left[\begin{array}{c} a_{t}[q]  \\ b_{t}[q]  \end{array}\right]$. Then for each single qubit gate $U_t^1[q]$, there is a term in the circuit Hamiltonian $H_{circuit}$ equal to 
\begin{equation*}
H_t^1[q]=[ C_{t}^{\dagger}-\lambda C_{t-1}^{\dagger}\,{U_t^1}^{\dagger} ] [ C_{t}-\lambda U_t^1 \,C_{t-1}],
\end{equation*}
where we have dropped the label $[q]$ for readability. This is a fermion hopping term for the $q$th fermion from site $t-1$ to $t$ and vice-versa, while $U_t^1$ acts on the internal spin degree of freedom. By including the onsite terms $C_t^{\dagger} C_t$ and $C_{t-1}^{\dagger} C_{t-1}$ one ensures that $H_t^1[q] \geq 0$. The parameter $\lambda \in [0,1]$ can tune the relative strength of the hopping, but we will take $\lambda=1$ for the rest of the paper.
In order for the circuit Hamiltonian to represent the action of a quantum circuit with some single qubit gates, we must require that the fermionic occupation number $N[q]=\sum_{t=0}^D n_{t}[q]=1$ with $n_t[q]\equiv a^{\dagger}_t[q]a_t[q]+b^{\dagger}_t[q]b_t[q]$, or that one qubit $q$ is represented by a single fermion present. If the original quantum circuit is universal, it will also involve CNOT gates (or controlled-U gates). The authors in \cite{MMC:groundstate} represent a CNOT gate between qubit $c$ (control) and $g$ (target) at time $t$ by the following two terms $H_t^{CNOT}[c,g]=H_t^I[c,g]+H_t^{NOT}[c,g]$ in the circuit Hamiltonian, i.e. 
%\begin{eqnarray}
%\lefteqn{H_t^I[c,g]= a_t^{\dagger}[c] a_t[c] \;n_t[g] +a_{t-1}^{\dagger}[c] a_{t-1}[c] \;n_{t-1}[g]} & \nonumber \\
%& -\left(a_t^{\dagger}[c] a_{t-1}[c] \left(a_t^{\dagger}[g] a_{t-1}[g]+b_t^{\dagger}[g] b_{t-1}[g] \right)+h.c.\right),
%\nonumber \\
%\lefteqn{H_t^{NOT}[c,g]= b_t^{\dagger}[c] b_t[c] n_t[g] +b_{t-1}^{\dagger}[c] b_{t-1}[c] n_{t-1}[g]} & \nonumber \\
%&-\left(b_t^{\dagger}[c] b_{t-1}[c] \left(a_t^{\dagger}[g] b_{t-1}[g]+b_t^{\dagger}[g] a_{t-1}[g]\right)+h.c.\right)
%\label{eq:CNOT}
%\end{eqnarray}
\begin{eqnarray}
H_t^I[c,g]=& a_t^{\dagger}[c] a_t[c] \;n_t[g] +a_{t-1}^{\dagger}[c] a_{t-1}[c] \;n_{t-1}[g]  \nonumber \\
& -\left(a_t^{\dagger}[c] a_{t-1}[c] \left(a_t^{\dagger}[g] a_{t-1}[g]+b_t^{\dagger}[g] b_{t-1}[g] \right)+h.c.\right),
\nonumber \\
H_t^{NOT}[c,g]=& b_t^{\dagger}[c] b_t[c] n_t[g] +b_{t-1}^{\dagger}[c] b_{t-1}[c] n_{t-1}[g] \nonumber \\
&-\left(b_t^{\dagger}[c] b_{t-1}[c] \left(a_t^{\dagger}[g] b_{t-1}[g]+b_t^{\dagger}[g] a_{t-1}[g]\right)+h.c.\right).
\label{eq:CNOT}
\end{eqnarray}
Note that for a general controlled-$U$ gate, we could take $H_t^{CU}[c,g]=H_t^I[c,g]+H_t^{U}[c,g]$ with the formal definition
\begin{eqnarray}
H_t^U[c,g]=& b_t^{\dagger}[c] b_t[c] \;n_t[g] +b_{t-1}^{\dagger}[c] b_{t-1}[c] \;n_{t-1}[g] \nonumber \\
&- \left(b_t^{\dagger}[c] b_{t-1}[c] \;C_t^{\dagger}[g] U C_{t-1}[g]+h.c.\right).
\nonumber
\label{eq:CU}
\end{eqnarray}
For such two-qubit gates, the fermions corresponding to qubits $c$ and $g$ both hop forward or backward and the internal spin-state of fermion $g$ is changed depending on the internal state of fermion $c$. If the original quantum circuit is 1-dimensional, then the circuit Hamiltonian describes a fairly natural interacting fermion system in 2D. It may thus be a physically attractive system for realizing quantum adiabatic computation \cite{MLM:qadiabatic} or quantum walks \cite{CGW:walk}. Note that these interactions preserve the condition that $\forall q,\;N[q]=1$. The authors in \cite{MLM:qadiabatic} propose to use the parameter $\lambda$ to adiabatically turn the dynamics of the terms $H_t^1[q]$ (and similarly $H_2^t[q]$) on.

First, we would like to note that this model of interacting fermions can be unitarily mapped onto the space-time circuit model introduced in Section \ref{sec:stcircuit} by the following steps \cite{thesis:breuckmann}. Instead of fermions, one can represent each qubit $q$ by a double line of $2(D+1)$ qubits as one can verify that the interactions remain local under a Jordan-Wigner transformation (note that the fermion hopping dynamics is that of nearest-neighbor coupled one-dimensional hopping). Then we unitarily switch the representation of the internal two-qubit state of the fermion at site $t$ from a `dual rail' representation to a representation in which the first qubit labels the clock and the second the current qubit state, i.e. we transform $\ket{01} \rightarrow \ket{10}$, $\ket{10} \rightarrow \ket{11}$, $\ket{00} \rightarrow \ket{00}$ and $\ket{11} \rightarrow \ket{01}$. The last input state $\ket{11}$ does not occur as $N[q]=1$. After these 2-qubit unitary transformations on all the qubits, we note that of the $2(D+1)$ qubits representing one qubit in the original circuit, $D$ qubits, out of $D+1$ qubits, are in the $\ket{0}$ state, while one qubit state has the current information. The other $D+1$ qubits represent the clock of the qubit as $\ket{t}=|0\rangle_1 |0\rangle_2 \ldots |0\rangle_{t} |1\rangle_{t+1} |0\rangle_{t+2} \ldots |0\rangle_{D+1}$. Note that the extra $D$ qubits in the $\ket{0}$ state can be unitarily transformed away, by moving swapping the information-containing qubit to the first qubit depending on the clock-register $\ket{t}$.

This clock representation is usually called a pulse clock, as opposed to a domain wall clock which was originally introduced in \cite{KSV:computation}. In our formulation of the circuit Hamiltonian we have not yet specified a particular clock realization; we discuss this in Section \ref{sec:clock}.

As the fermionic circuit Hamiltonian in the sector $N[q]=1$ for all qubits $q$, is unitarily related to the circuit Hamiltonian in Section \ref{sec:stcircuit}, the spectrum of the Hamiltonians is the same. In \cite{MLM:qadiabatic, MMC:scaling} the authors provide bounds on the gap above the ground-space. In \cite{MLM:qadiabatic} a penalty term $H_{causal}$ is added to $H_{circuit}$ which ensures that invalid configuration have at least some constant energy, see Eq.~(\ref{eq:causmlm}) in Section \ref{sec:QMA}. 

The authors claim that the lowest nonzero eigenvalue of $H_{circuit}$ in the space of valid time configurations is $\Omega(S^{-4})$ where $S$ is the size of the quantum circuit. The proof of this claim is however not contained in \cite{MLM:qadiabatic}, but the authors refer back to section C in \cite{MMC:scaling} where this result seems to be claimed for any quantum circuit consisting of single qubit and two-qubit gates. However, the arguments in Section C in \cite{MMC:scaling} make no reference to having to exclude invalid time-configurations which can easily be constructed to have zero energy. We believe that the gap analysis in these papers misses several essential and interesting aspects of the space-time circuit-to-Hamiltonian construction and warrants a more thorough mathematical investigation. This is what we set out to do in this paper.

\subsection{Space-Time Circuit-to-Hamiltonian Construction with Circular Time}
\label{sec:STcirctime}
\begin{figure}[htb]
\centering
\includegraphics[width=0.2\hsize]{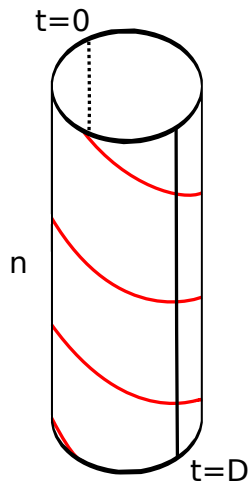} 
\caption{Space-Time Cylinder with circumference $2D$ and length $n$ with $n=6D$, based on quantum circuit in Fig.~(\ref{fig2}b). We identify the top and bottom of cylinder (periodic boundaries in space) to make a torus. The red line represents a zero energy time configuration, a closed time-loop. Such zero energy loops can be constructed whenever $n=2kD$ with integer $k$.}
\label{fig3}
\end{figure}

The construction in Sec.~\ref{sec:stcircuit} gets modified when the clock registers represent a circular time.
For {\em each} qubit $q$ in the original circuit, we define an individual clock register $\ket{t}_q$ with $t \in Z_{2D}$. For simplicity, we again assume that the quantum circuit only contains two-qubit gates. 
One possible construction is to take $H_{circuit}=\sum_{t=1}^{2D} H_t$ where $H_t$ is a sum over terms $H^2_t[q,p]$ corresponding to all the gates which occur in parallel at time-step $t$ in the original circuit, i.e. Eq.~(\ref{eq:2q}) for $t \in [1,D]$. For $t\in [D+1,2D]$ we take terms corresponding to the inverses of all the gates which occur at time-step $2D-t+1$. However, if we apply this to the circuit in Fig.~(\ref{fig2})(b), we loose the alternating structure of the quantum circuit at times $t=0$ and $t=D$. We can simply avoid this by assuming that in the last time-step of the circuit only $I$ gates are performed on all qubits. Instead of undoing this gate in the next time-step at $t=D+1$, we 'undo' it in the last time-step $t=2D$. 
Thus the terms $H_t$ for $t \in [1,D]$ correspond again to the original two-qubit gates. The terms $H_t$ with $t \in [D+1,2D-1]$ correspond to the inverses of gates happening at time-steps $2D-t$ and the last term $H_{2D}$ corresponds to the (trivial $I$) gates happening at time $t=D$ in the original quantum circuit. In this way, we can wrap the alternating gate structure around a cyclinder, Fig.~(\ref{fig3}).

What are the zero energy states for such ciruit Hamiltonian? We will have to redefine what it means for time-configurations to be invalid as compared to Section \ref{sec:stcircuit} as there is no notion of `after' or `before' a certain time when time is circular. A two-qubit gate $U^2[q,p]$ occurring at time $t$ in the quantum circuit gets mapped onto two terms in $H_{circuit}$ in general. The gate specifies two complementary time intervals between the two gate-terms, $I_t$ and $I_t^c$ with $I_t \cup I_t^c =Z_{2D}$. For example, for the unraveling choice above, all gates at timesteps $t \in [1,D)$, the intervals are $I_t=[t,2D-t-1]$ and $I_t^c=[2D-t,t-1]$ and for the $I$-gates at $t=D$, the intervals are $[D,2D-1]$ and $[0,D-1]$. A time-configuration $t_1,\ldots, t_n$ is called invalid if there exists at least one pair of such qubits $(q,p)$ interacting at time $t$ in the original circuit, for which either $(t_q \in I_t) \wedge (t_p \in I_t^c)$ or $(t_p \in I_t) \wedge (t_q \in I_t^c)$. 

%NB removed l.l.
%Next, we consider zero energy light-like configurations. 
We consider valid zero energy configurations. If we impose periodic boundaries conditions in space and take circular time with $n=2 k D$ with integer $k=1,2,\ldots$, one can construct zero-energy configurations, see Fig.~(\ref{fig3}). The configuration with (even) $n=2k D$ makes a homologically nontrivial loop around the torus in both directions (one always makes a nontrivial loop around the space-direction). For $n < 2D$ and two-qubit gates throughout the quantum circuit, we note that it is not possible to have such zero-energy configurations. 
%NPB added "homologically"

\section{Gap of the Circuit Hamiltonian}
\label{sec:gap}

In this section we will do the technical work of lowerbounding the gap of the circuit Hamiltonian for one-dimensional quantum circuits with closed boundary conditions in space, Fig.~(\ref{fig2})(b), in which the circuit Hamiltonian is constructed using circular time as in Sec.~\ref{sec:STcirctime}. We start with some observations which hold for more general quantum circuits. We consider the gap of the circuit Hamiltonian in the space of valid time-configurations. Such valid time configurations will be denoted as $\ket{{\bf t}}$. We can associate a graph and its Laplacian with the circuit Hamiltonian on this valid subspace spanned by $\ket{{\bf t}}$. Let $G=(V,E)$ be a graph with vertices ${\bf t} \in V$ representing valid time-configurations and let $E$ be the set of undirected edges of the graph.  There exists an edge $e=({\bf t},{\bf t'}) \in E$ between valid time-configurations ${\bf t} \neq {\bf t'}$ iff 
\begin{equation*}
\bra{\bf t} H_{circuit} \ket{\bf t'}=-V({\bf t} \leftarrow {\bf t'}) \neq 0, \nonumber
\end{equation*}
for some unitary $V({\bf t} \leftarrow {\bf t'})$, i.e. $V({\bf t} \leftarrow {\bf t'})$ is the particular single-qubit or two-qubit gate of the quantum circuit which connects ${\bf t'}$ to ${\bf t}$. The Laplacian of the graph underlying the circuit Hamiltonian is defined as
\begin{equation*}
L(G)_{{\bf t},{\bf t'}}=\left\{\begin{array}{ll}
\mbox{deg}({\bf t}), & {\bf t}={\bf t'} \\ 
-1, & ({\bf t},{\bf t'}) \in E \\
0  &  else.
\end{array}\right.
\end{equation*}
Note that one can write $L(G)=D(G)-A(G)$ with diagonal degree matrix $D(G)$ and adjacency matrix $A(G)$. 

If $G$ is a {\em connected} graph then by some number of applications of $H_{circuit}$ one can get from any valid time-configuration to any other one. We will be only interested in {\em connected graphs}: this precludes the existence of
%NB remove ll reference 
%light-like configurations, or more generally 
disconnected clusters of valid time-configurations. It may be clear that for the one-dimensional quantum circuit with two-qubit gates throughout with a circular time and $2D > n$, Fig.~(\ref{fig2}b), $H_{circuit}$ corresponds to a connected graph. For a connected graph, one can always construct a path from the `origin' time-configuration ${\bf t}=(0,0,\ldots 0)={\bf 0}$ to any other ${\bf t}$. It may also be clear that there is a {\em unique} unitary transformation $V({\bf t}\leftarrow {\bf 0})=V({\bf t} \leftarrow {\bf t}_m)\ldots V({\bf t}_2 \leftarrow {\bf t}_1)V({\bf t}_1 \leftarrow {\bf 0})$ which one can associate with such a path (of length $m+1$) \footnote{Note that the path may not be unique as the order in which the gates are executed is not unique, but the induced unitary transformation will nonetheless be unique.}. Using this composite unitary transformation $V({\bf t} \leftarrow {\bf 0})$ we can transform away the dependence of $H_{circuit}$ on the particular unitary gates. That is, let \begin{equation}
W=\sum_{valid\, {\bf t}} V({\bf t} \leftarrow {\bf 0}) \ket{\bf t} \bra{\bf t},
\label{eq:defW}
\end{equation} then
\begin{equation}
W^{\dagger} H_{circuit}(\{U\},G) W=H_{circuit}(\{U=I\},G)=\sum_{{\bf t},{\bf t'}} L(G)_{{\bf t},{\bf t'}} \ket{\bf t}\bra{\bf t'}.
\label{eq:u_equiv}
\end{equation}

The standard Feynman-Kitaev construction is a simple example of this in which the underlying graph is a one-dimensional line or circle and is thus connected. The space-time circuit-to-Hamiltonian construction generalizes this to high-dimensional graphs whose vertices are no longer points but strings (for one-dimensional circuits) or membranes (for two-dimensional quantum circuits) etc.

From the spectral theory of Laplacians on graphs \cite{mohar:lap}, one can get some standard results, e.g. 

\begin{prop}The lowest eigenvalue of the Laplacian of a connected graph $G=(V,E)$ is zero and corresponds to a unique vector which is the uniform superposition over all vertices. 
\end{prop} 

This directly implies that for circuit Hamiltonians with underlying connected graph $G=(V,E)$, the unique ground-state in the space of valid time-configurations is the history state
\begin{equation*}
\ket{\psi_{history}}=\frac{1}{\sqrt{|V|}} \sum_{valid\;{\bf t}} V({\bf t} \leftarrow {\bf 0}) \ket{\xi} \otimes \ket{\bf t}, \forall \xi.
\end{equation*}

The second smallest eigenvalue of the Laplacian of a graph (and thus the gap of the circuit Hamiltonian) is called the algebraic connectivity. Various techniques have been developed to bound this eigenvalue \cite{mohar:lap}, in particular using the theory of random walks on graphs and their mixing times.

For the one-dimensional quantum circuit in Fig.~(\ref{fig2})(b), with the circular-time $H_{circuit}$, the graph is translationally-invariant in the `time direction'. Due to the periodic boundaries conditions in space, the valid time-configurations corresponds to strings which wind around the torus, see Fig. \ref{fig3}. This model is identical to the model considered in \cite{BS:randomwalk_on_randomwalk}. Our question, namely bounding the mixing time of the process of diffusion of a closed string, is slightly different from the problem solved in that paper.  The problem of diffusion of a domain wall (of an ferromagnetic Ising model at $T=0$ where the Ising spin $+1$ or $-1$ represents whether a gate has been done or not) has also been considered in the condensed-matter literature, see e.g. \cite{PRL:growth, SKR:fate}.

\subsection{One-dimensional Quantum Circuits: FM Heisenberg Model Coupled to a Counter}
\label{sec:FMHeis}

% BMT even n

\begin{figure}[ht]
   \centering
      \subfloat[]{\includegraphics[scale=0.3]{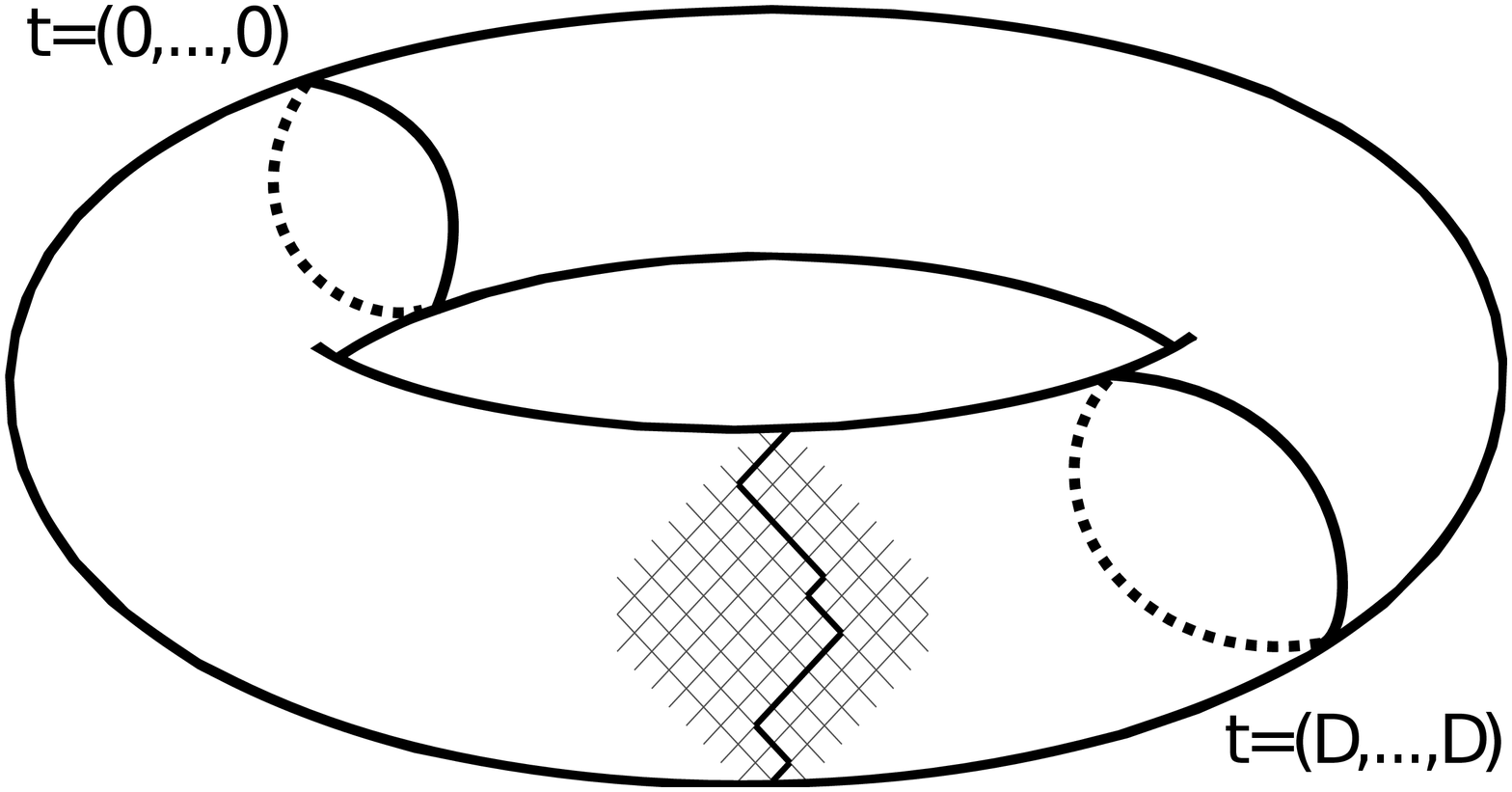} }\qquad%NPB space between subfloats
      \subfloat[]{\includegraphics[scale=0.1]{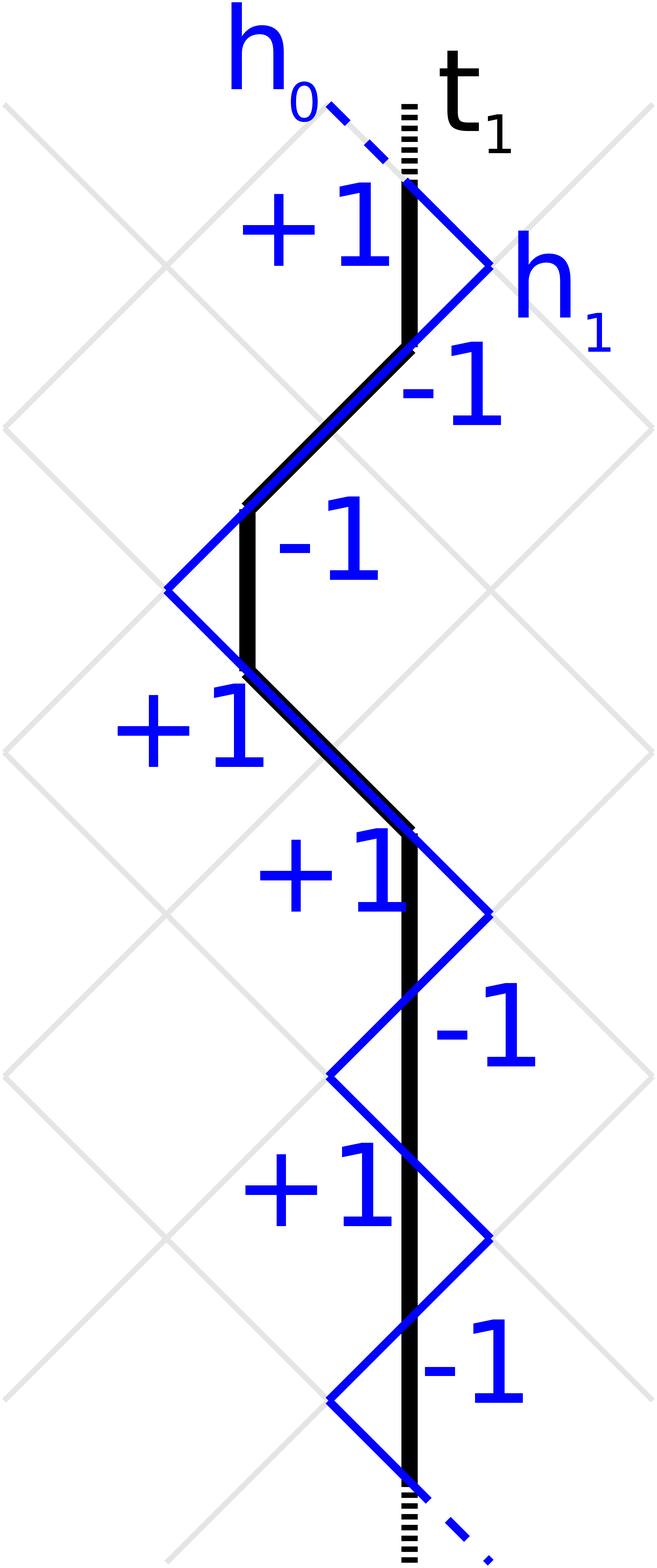} }
   \caption{(a) The valid time configurations of the quantum circuit in Fig.~(\ref{fig2})(b), using the circular-time construction, can be represented as a single string which winds around the torus. The dynamics of the circuit Hamiltonian corresponds to diffusion of this string. The square plaquettes represent the gates and the string forms the boundary of the gates that have already been executed. (b) Relabeling of the string variables using the boundary point $h_0$ which is next to the time $t_1$ of qubit $1$ and the variables $x_i$ with $(-1)^{x_i}=\pm 1$ which indicate whether the string continues left or right.}
\label{fig4}
\end{figure}

%BMT some clarifications

We start with a convenient relabeling of the valid time-configurations ${\bf t}$ as  $(\tau, x)$ where $\tau \in Z_{D}$ and bitstring $x=x_1, x_2, \ldots, x_n$ in the following manner. Let $t_1$ be the time of one designated qubit, say, qubit $1$. We assume as in Fig.~(\ref{fig2})(b) that the first gate on qubit $1$ is between qubits $1$ and $2$. Let $h_0=t_1+\frac{1}{2}$ if $t_1$ is even and $h_0=t_1-\frac{1}{2}$ if $t_1$ is odd so that $h_0$ takes on values $\frac{1}{2}+2 \tau$ with $\tau \in Z_{D}$, see Fig.~(\ref{fig4})(b). Each valid time-configuration can be associated with the half integers $h_0,h_1, \ldots, h_{n-1}$ ($h_n=h_0$) which are defined at the vertices of the square plaquettes in Fig.~(\ref{fig4})(b) such that $(-1)^{x_i}=h_{i}-h_{i-1}$. It is clear from the Figure that a string ${\bf t}$ is equivalent to $(h_0, \ldots, h_{n-1})$ which is equivalent to $(\tau, x_1,\ldots,x_n)$ with $x_i=0,1$. Essentially, we are just reparametrizing the string ${\bf t}$ in terms of a point through which the string crosses and deviations from this point which of course fully determines the position of the string. Note that we explicitly break the translation symmetry between the qubits with this parametrization. It is important to note that the periodic boundary conditions in space imply that $\sum_{i=1}^n (-1)^{x_i}=0$ or $\sum_{i=1}^n Z_i=0$, i.e. an equal number of `spins' are up or down. 

This relabeling also immediately gives us the number of vertices in the graph $G=(V,E)$ as $|V|=D {n \choose n/2}$. We consider the action of the circuit Hamiltonian (omitting the unitary gates due to Eq.~(\ref{eq:u_equiv})) in this relabeled basis. Note that terms in $H_{circuit}$ which correspond to gates between qubits $1$ and $n$ act on $h_0$ and the `spin' states $x_1$ and $x_n$. By such term $h_0$ can be mapped onto $h_0 \pm 2$ or the counter variable $\tau$ to $\tau \pm 1$.

Terms which correspond to gates between the other qubits do not act on the counter $\tau$ but {\em only} on the spin states. For adjacent variables $\ket{x_i=0, x_{i+1}=1} \leftrightarrow \ket{x_i=1,x_{i+1}=0}$ while $\ket{x_i=1,x_{i+1}=1}$ or $\ket{x_i=0,x_{i+1}=0}$ are left unchanged. The dynamics of the internal variables $x$ corresponds to that of the isotropic ferromagnetic spin-1/2 Heisenberg model with the condition $\sum_{i=1}^n Z_i=0$. More precisely, the circuit Hamiltonian (in the valid time-config. subspace) is unitarily equivalent to 
\begin{eqnarray}
\tilde{H}_{circuit}=&\sum_{i=1}^{n-1} (\sigma_i^+ \sigma_i^- \sigma_{i+1}^- \sigma_{i+1}^++\sigma_i^- \sigma_i^+ \sigma_{i+1}^+ \sigma_{i+1}^-) -
\sum_{i=1}^{n-1} (\sigma_i^+ \sigma_{i+1}^-+h.c) \nonumber \\
&+\left(\sigma_n^+ \sigma_n^- \sigma_{1}^- \sigma_{1}^++\!\sigma_n^- \sigma_n^+ \sigma_{1}^+ \sigma_{1}^- \right) 
-\!(\sigma_1^-\sigma_n^+ \sum_{\tau=0}^{D-1} \ket{\tau-1}\bra{\tau}+\!h.c.)
\label{eq:sameH}
\end{eqnarray}
One can verify this form of the Hamiltonian by inspecting the matrix elements $\bra{{\bf t}} H_{circuit}(U=I) \ket{{\bf t'}}=L_{{\bf t},{\bf t'}}$, Eq.~(\ref{eq:u_equiv}), and representing ${\bf t}$ in terms of $\ket{\tau,x}$. The off-diagonal terms with negative sign directly come from minus the adjacency matrix, $-A_{{\bf t},{\bf t'}}$, while the positive diagonal terms arise from the diagonal degree matrix $D_{{\bf t},{\bf t'}}$.

The eigenstates of $\tilde{H}_{circuit}$ with respect to the counter variable $\tau$ are simple plane-waves, i.e.  
\begin{eqnarray}
\ket{\psi_k}=\frac{1}{\sqrt{D}} \sum_{\tau=0}^{D-1} e^{2 \pi i k \tau /D} \ket{\tau}, \;\;k=0,\ldots D-1, \nonumber \\ 
\tilde{H}_{circuit} \ket{\psi_k}\otimes \ket{\phi} =\ket{\psi_k} \otimes H(k) \ket{\phi},
\label{eq:hk}
\end{eqnarray}
%NB added \ket{\phi}
where $\ket{\phi}$ is any state of the spins. Using $\sigma_i^+ \sigma_{i+1}^-+h.c=\frac{1}{2}(X_i X_{i+1}+Y_i Y_{i+1})$ and $\sigma_i^{+} \sigma_i^{-}=\frac{1}{2}(I-Z_i)$ we have 
\begin{equation}
H(k)=\frac{n-1}{2}-\frac{1}{2}\sum_{i=1}^{n-1} (X_{i} X_{i+1}+Y_i Y_{i+1}+ Z_i Z_{i+1})+\Delta(k),
\label{eq:splitheis}
\end{equation}
with
\begin{equation}
\Delta(k)=\frac{1}{2}(1-Z_1 Z_n)-\sigma_1^- \sigma_n^+ e^{2 \pi i k/D}-\sigma_1^+ \sigma_n^- e^{-2 \pi i k/D}\geq 0.
\label{eq:boundterm}
\end{equation}

The eigenstates (and eigenvalues) of $\tilde{H}_{circuit}$ are thus the eigenstates of $H(k)$ in tensorproduct with the plane-wave states $\ket{\psi_k}$. $H(k=0)$ is the ferromagnetic (spin-$\frac{1}{2}$) Heisenberg chain with periodic boundary conditions (in the sector with $\sum_i Z_i=0$), i.e. 
\begin{equation}
H(k=0)=\frac{n}{2}-\frac{1}{2}\sum_{i=1}^n (X_{i} X_{i+1}+Y_i Y_{i+1}+ Z_i Z_{i+1}) \geq 0.
\label{eq:closeheis}
\end{equation}
This model can be analyzed using the Bethe ansatz, see e.g. \cite{thesis:ng}. Note that the condition $\sum_i Z_i=0$ is not the usual one studied in physics: one can interpret it as there being $n/2$ particles (out of $n$) which by the dynamics of $H(k)$ can interchange positions on a circle. The model $H(k \neq 0)$ corresponds to a ferromagnetic Heisenberg chain with a {\em partially twisted boundary}.  It may be possible to obtain the full spectrum of the partially-twisted Heisenberg chain $H(k)$ with a Bethe ansatz, but here we focus on determining the lowest eigenvalues.

The unique groundstate of $\tilde{H}_{circuit}$ is the zero energy groundstate of $H(k=0)$, the state $\frac{1}{\sqrt{D {n \choose n/2}}}\sum_{\tau=0}^{D-1} \sum_{x: \sum (-1)^{x_i}=0}\ket{\tau, x}$. 

The gap of the ferromagnetic Heisenberg chain $H(k=0)$ for $n$ spins with $\sum_i Z_i=0$ has been lowerbounded previously, see Theorem \ref{thm:DS} in Section \ref{sec:openheis}.
In order to lowerbound the gap of $\tilde{H}_{circuit}$, we also need to lowerbound the groundstate energies for any $H(k\neq 0)$. Let us outline the remainder of our proof. We have $H(k)=A+B$ where $A$ is the ferromagnetic Heisenberg chain with {\em open boundaries}, i.e. let 
\begin{equation}
A \equiv \frac{n-1}{2}-\frac{1}{2}\sum_{i=1}^{n-1} (X_{i} X_{i+1}+Y_i Y_{i+1}+ Z_i Z_{i+1}) \geq 0
\label{eq:openheis}
\end{equation}
and $B \equiv \Delta(k\neq 0)$. We will invoke the following lemma 

\begin{lem}[Kitaev\cite{KSV:computation}]
Let $A \geq 0$ and $B \geq 0$ and let $\ker(A)/ \ker(B)$ be their respective nullspaces, where $\ker(A)\cap \ker(B)=\lbrace 0 \rbrace$. Let $\lambda_1(A)$ ($\lambda_1(B)$) be the smallest nonzero eigenvalue of $A$ ($B$). Then
\begin{equation*}
A+B \geq \min(\lambda_1(A),\lambda_1(B)) \cdot  (1-\cos(\theta)). \nonumber
\end{equation*}
with $\cos(\theta)=\max_{\psi_B \in \ker(B), \psi_A \in \ker(A)}|\langle \psi_A | \psi_B \rangle|$.
\label{lem:kitaev}
\end{lem}
 
Thus if we can bound the gap of $A$ (see Eq.~(\ref{eq:openb}) in Section \ref{sec:openheis}) and bound the gap of the boundary term $\Delta(k \neq 0)$ (this is simple as it involves two qubits) and bound the angle between the two null-spaces $\ker(A)$ and $\ker(B)$ (see Lemma \ref{lem:angle}), we can obtain a lowerbound on the smallest eigenvalue of $H(k \neq 0)$. Together with the lowerbound on the gap of $H(k=0)$, Theorem \ref{thm:DS}, this will prove the following result:

\begin{thm}
The smallest non-zero eigenvalue $\lambda_1$ of the  Hamiltonian $H_{circuit}$ of a one-dimensional, depth $D>\frac{n}{2}$, quantum circuit on $n$ qubits  in the space of valid time-configurations, is bounded as 
\begin{equation}
\lambda_1(H_{circuit})=\lambda_1(\tilde{H}_{circuit}) \geq \frac{\pi^4}{4D^2(n-1)n}+O\left(\frac{1}{n^4 D^2}\right).
\label{eq:gapbound}
\end{equation}
\label{thm:lb}
\end{thm}

{\em Proof}:
As we argued before, the spectrum of $H_{circuit}$ is the same as the spectrum of $\tilde{H}_{circuit}$ which in turn is the same as the union of spectra of $H(k)$ for all $k$ due to Eq.~(\ref{eq:hk}). Theorem \ref{thm:DS} shows that $\lambda_1(H(k=0))=\Omega(\frac{1}{n^2})$, but $H(k \neq 0)$ may have lower nonzero eigenvalues. We invoke Lemma
\ref{lem:kitaev}. We have $\lambda_1(B) \geq 2$ by direct calculation and we use Eq.~(\ref{eq:openb}) to lowerbound $\lambda_1(A)$. The angle between the null-spaces $\ker(A)$ and $\ker(B)$ is given in Lemma \ref{lem:angle}. This results in Eq.~(\ref{eq:gapbound}). $\Box$

\begin{lem}[Angle between Subspaces] Let $A$ be the open-boundary Heisenberg chain defined in Eq.(\ref{eq:openheis}) and let $B$ be the boundary term $B=\Delta(k \neq 0)$ defined in Eq.~(\ref{eq:boundterm}). Furthermore, let $\mathcal{H}$ be the subspace where $\sum_{i} Z_{i}=0$ and $\cos(\theta)=\max_{\psi_B \in \ker(B)\cap \mathcal{H}, \psi_A \in \ker(A)\cap \mathcal{H}}|\langle \psi_A | \psi_B \rangle|$. Then 
\begin{equation*}
1-\cos(\theta) \geq \frac{\pi^2 n}{4 D^2 (n-1)}+O\left(\frac{1}{D^4}\right).
\end{equation*}
\label{lem:angle}
\end{lem}

{\em Proof}: The groundstate $\ket{\psi_A^0}={n \choose n/2}^{-1/2} \sum_{x \colon \sum_i (-1)^{x_i}=0} \ket{x}$ of $A$ is unique, see also Section \ref{sec:openheis}.  Thus we consider
\begin{equation*}
1-\cos(\theta)=\min_{\psi_B \in {\rm Ker}(B)} \left(1-\sqrt{F(\psi_A^0,\psi_B)} \right),
\end{equation*}
with the fidelity $F(\sigma, \rho)=\left({\rm Tr} \sqrt{\rho^{1/2} \sigma \rho^{1/2}}\right)^2$ for two arbitrary density matrices $\sigma$ and $\rho$. We can use the monotonicity of fidelity under taking partial traces, i.e. $ F(\rho_A^0,\rho_B) \geq F(\psi_A^0,\psi_B)$ \cite{book:nielsen&chuang} for the reduced density matrices $\rho_A^0$ and $\rho_B(k)$ for qubits $1$ and $n$. The reduced density matrix of $\psi_A^0$ equals 
\begin{equation*}
\rho_A^0=\frac{n-2}{4(n-1)}\left(\ket{00}\bra{00}+\ket{11}\bra{11}\right)+\frac{n}{2(n-1)}\ket{\eta_0}\bra{\eta_0},
\end{equation*}
with $\ket{\eta_0}=\frac{1}{\sqrt{2}}(\ket{01}+\ket{10})$. The space $\ker{B}$ is spanned by vectors of the form $\ket{00}\otimes\ket{\psi_{00}}$,$\ket{11}\otimes\ket{\psi_{11}}$ and $\ket{\eta_k}\otimes\ket{\psi_{\eta_k}}$ with $\ket{\eta_k}=\frac{1}{\sqrt{2}} (\ket{01}+e^{-2\pi i k/D} \ket{10})$. Here $\ket{\psi_{00}}, \ket{\psi_{11}},\ket{\psi_{\eta_k}}$ are orthogonal as they contain a different number of particles (remember $\sum_i Z_i=0$).  As the states in the nullspace of $B$ are not fully symmetric under all permutations of particles, the null-spaces of $A$ and $B$ have zero intersection.
A reduced density matrix $\rho_B(k)$ can thus be parametrized as  
\begin{equation*}
\rho_B(k)=|\alpha|^2 \ket{00}\bra{00}+|\beta|^2\ket{11}\bra{11}+|\gamma|^2 \ket{\eta_k}\bra{\eta_k}, 
\end{equation*}
with $|\alpha|^2+|\beta|^2+|\gamma|^2=1$, so that
\begin{equation*}
{\rm Tr} \left(\rho_B^{1/2}(k) \,\rho_A^0 \,\rho_B^{1/2}(k)\right)^{1/2}=(|\alpha|+|\beta|)\sqrt{\frac{n-2}{4(n-1)}}+|\gamma|\sqrt{\frac{n}{2(n-1)}}|\langle \eta_0| \eta_k\rangle|. \nonumber
\end{equation*}
Using the Cauchy-Schwartz inequality and $|\bra{\eta_0} \eta_k \rangle|^2=\frac{1+\cos(2 \pi k/D)}{2}$ we can upperbound 
\begin{equation*}
\sqrt{F(\rho_A^0,\rho_B(k))} \leq \sqrt{\frac{2(n-2)}{4(n-1)}+\frac{n(1+\cos(2\pi k/D))}{4(n-1)}}.
\end{equation*}
This fidelity is clearly maximized for the lowest non-zero momentum $k=1$ (or $k=D-1$) so that, using the Taylor expansion for the cosine and square-root, we can bound
\begin{equation*}
\sqrt{F(\psi_A^0,\psi_B)} \leq 1-\frac{\pi^2 n}{4 D^2 (n-1)}+O\left(\frac{1}{D^4}\right).
\end{equation*}
$\Box$

\subsubsection{Heisenberg Chain With (Open) Boundaries: connection with Markov chains}
\label{sec:openheis}

The ferromagnetic Heisenberg chain Hamiltonian with closed or open boundaries commutes with each of the su(2) spin operators $\vec{S}=(S_x, S_y,S_z)$ where $S_{\alpha}=\frac{1}{2}\sum_{i=1}^n \sigma_{\alpha}^i$ with $\sigma^i=(X_i,Y_i,Z_i)$. Using the total spin operator $S^2=\vec{S} \cdot \vec{S}$ which commutes with all $S_{\alpha}$, one can thus label the eigenstates by the quantum numbers $\ket{s,m}$, $m=-s,\ldots,s$ with $S_z \ket{s,m}=m \ket{s,m}$ and $S^2 \ket{m,s}=s(s+1) \ket{m,s}$. 

We are interested in the sector where $S_z=\frac{1}{2}\sum_i Z_i$ has eigenvalue $m=0$. The groundstate in this sector is degenerate with the overall ground-state which can easily be seen as follows. As the Heisenberg Hamiltonian $H(k=0)$ (periodic boundaries) or $A$ (open boundaries) is positive semidefinite, the state $\ket{000 \ldots 0}$ is a zero-energy groundstate with $m=n/2$. Using the lowering operator $S_-=S_x-i S_y$ which acts as $S_-\ket{s,m} \propto \ket{s,m-1}$ and noting that the lowering operator $S_-$ commutes with the isotropic Heisenberg Hamiltonian one can reach an eigenstate with zero-energy in the $m=0$ sector. This implies that the gap of the Heisenberg model in the $m=0$ sector can be lowerbounded by the gap of the Heisenberg model without specifying any sector. For open boundary conditions, Ref.~\cite{TN:gapXXX} lowerbounds this gap as
\begin{equation}
\lambda_1(A) \geq 2(1-\cos(\pi/n))=\Omega\left(\frac{1}{n^2}\right).
\label{eq:openb}
\end{equation}

It is expected that similar results hold for the gap of the Heisenberg model with periodic boundaries, but we will invoke a nice and well-known connection to the theory of Markov chains.  We use the relation between the Heisenberg model and a particle interchange model, see e.g. \cite{thesis:ng}. Let $P_{i,i+1}$ be a transposition (permutation) of particles at $i$ and $i+1$, i.e. $P_{i,i+1} \ket{01}_{i,i+1}= \ket{10}_{i,i+1}$, $P_{i,i+1}\ket{10}_{i,i+1}=\ket{01}_{i,i+1}$ and $P_{i,i+1}\ket{11}_{i,i+1}=\ket{11}_{i,i+1}$ and $P_{i,i+1} \ket{00}_{i,i+1}=\ket{00}_{i.i+1}$. We can define the symmetric, stochastic Markov matrix $P(x,y)=\frac{1}{n} \sum_{i=1}^n \bra{y} P_{i,i+1}\ket{x}$ on the space of bitstrings $\ket{x}$ with $\sum_{i} (-1)^{x_i}=1$, or the space with $n/2$ particles (out of $n$). 
The Hamiltonian in Eq.~(\ref{eq:closeheis}) can then be written as $H(k=0)=n-\sum_{i=1}^n P_{i,i+1}$ or $\bra{y} H(k=0) \ket{x}=n(\delta_{xy}-P(x,y))$.

The Markov process given by $P(x,y)$ is reversible, irreducible and aperiodic. Thus $P$ has a unique fixed point $\pi(x)={n \choose n/2}^{-1}$ (see e.g. \cite{book:LPW}). The second largest eigenvalue of $P$ determines the smallest non-zero eigenvalue of the Heisenberg chain with a closed boundary. This second largest eigenvalue of $P$ has previously been bounded, i.e.

\begin{thm}[Theorem 3.1 in \cite{DS:comparison}, see also \cite{thesis:ng}]
Let $P$ be the reversible, irreducible Markov chain defined above with eigenvalues $\beta_0=1 > \beta_1 > \beta_2 \geq \ldots$. Then the second largest eigenvalue of $P$ is 
\begin{equation*}
\beta_1 \leq 1- \frac{12}{(n+1)(n/2+1)n},
\end{equation*}
which directly implies that
\begin{equation*}
\lambda_1(H(k=0)) \geq \frac{12}{(n+1)(n/2+1)}.
\end{equation*}
\label{thm:DS}
\end{thm}

\section{Application to QMA and Quantum Adiabatic Computation}

\subsection{QMA}
\label{sec:QMA}

As the general local Hamiltonian problem is contained in QMA \cite{KSV:computation},  it is the second part of the QMA-completeness which concerns us here. We construct a map from any class of problems $L=L_{yes} \cup L_{no}$ in ${\rm QMA}$ to a Hamiltonian, using the space-time construction, such that: 
\begin{itemize}
\item if $x \in L_{yes}$, then the Hamiltonian $H(x)$ has eigenvalue lower than or equal to some $a$, see Sec.~\ref{sec:YES}.
\item if $x \in L_{no}$, then all eigenvalues of the Hamiltonian are larger than or equal to $b$ where $|a-b| \geq \frac{1}{\rm poly(n)}$, see Sec.~\ref{sec:NO}.
\end{itemize}

A property that any promise problem $L$ in QMA possesses is the existence of the verification circuits $C_n$ with the properties in Definition \ref{def:QMA}. The quantum circuit $C_n$ takes as input the unspecified quantum proof $\ket{\xi}$ provided by Merlin and some initial input qubits in a set $S_{in}$ set to $\ket{0}$ or $\ket{1}$ with $|S_{in}|=m < n$. The instance $x$ is also part of this input set of qubits. Whether qubits in $S_{in}$ are set to 0 or 1 plays no role in the proof, so for notational simplicity we require the qubits in $S_{in}$ to be $\ket{0}$.

W.l.o.g. we can take the verification circuit to be of the form, Fig.~(\ref{fig2}), as such one-dimensional quantum circuits with only two-qubit gates are universal. The circuit acts on $n$ qubits and has depth $D$ which is a some polynomial in $n$. Let $q_{out}$ be the output qubit of the circuit $C_n$, so that $\prob{\left[C_n(x,\xi)=1\right]}=\prob{[q_{out}=1]}$.

For every qubit in the quantum circuit, one can define a past causal cone of qubits, namely those qubits which could have influenced the state of that qubit at the end of the computation. It is important to note that we may assume w.l.o.g. that the qubits in the set $S_{in}$ are in the past causal cone of the output qubit $q_{out}$. If they are not, then these qubits are not needed to produce this output so we could omit them.  The Hamiltonian which corresponds to a verification circuit is 
\begin{equation}
H=H_{circuit}+H_{in}+H_{out}+H_{causal}
\label{eq:defH}
\end{equation}
where $H_{circuit}$ is the space-time circuit Hamiltonian of the verification circuit in Fig.~(\ref{fig2})(b) with circular time.
Recall that we have shown that the unique zero energy ground-state (space) of this $H_{circuit}$ is of the form
\begin{eqnarray}
\ket{\psi_{history}}=\frac{1}{\sqrt{D {n \choose n/2}}} \sum_{valid\;{\bf t}} V({\bf t} \leftarrow {\bf 0}) \ket{\phi_{in}} \otimes \ket{\bf t}, \nonumber \\ \ket{\phi_{in}}=\sum_{y \in \{0,1\}^{m}} \alpha_y \ket{\xi_y}\ket{y}_{S_{in}}.
\label{eq:history}
\end{eqnarray}
Here $y$ are the input-qubits in $S_{in}$ and $\ket{\xi_y}$ is a general input state of the other qubits. One makes the following choice for $H_{in}$ and $H_{out}$:
\begin{eqnarray}
H_{in}&=\sum_{p \in S_{in}} \ket{1}\bra{1}_{p}\otimes \ket{t=0}\bra{t=0}_p , \nonumber \\ H_{out}&=\ket{0}\bra{0}_{q_{out}} \otimes \ket{t=D}\bra{t=D}_{q_{out}}. 
\label{eq:hin}
\end{eqnarray}
The term $H_{causal}$ is a penalty term for invalid time-configurations. It is a sum of terms, one for each two-qubit gate in the original quantum circuit. Let there be a gate acting at time $t$ on qubits $[q,p]$ in the original quantum circuit. Let $\Pi(t_q \in I_t)=\sum_{s \in I_t} \ket{s}\bra{s}_{q}$ where the interval $I_t$ (and $I_t^c$) were defined in Section \ref{sec:STcirctime}. Such projector acts on the time register of qubit $q$ and has eigenvalue $1$ if $t_q \in I_t$ (and 0 otherwise). The penalty term corresponding to this gate equals
\begin{equation}
H_{causal}([q,p],t)=\Pi(t_q \in I_t) \Pi(t_p \in I_t^c)+\Pi(t_p \in I_t) \Pi(t_q \in I_t^c).
\label{eq:caus}
\end{equation}
%NB added reference to local version of H_caus
$H_{causal}$ commutes with $H_{in}$ and $H_{out}$ as all terms are diagonal in the same basis. Note that $H_{causal}$ as defined here is not local; we will address this point in section \ref{sec:QMA2Dfermions}. Each term $H_{causal}([q,p],t)$ commutes with $H_{circuit}$ as follows. First of all, $H_{causal}([q,p],t)$  commutes with the two terms which represent the gate $U_t^2[q,p]$ in the circuit Hamiltonian, as $H_{causal}([q,p],t)H_t^2[q,p]=0$ etc. It obviously commutes with any $H_t^2[q',p']$ with $q' \neq q$ and $p'\neq p$. Lastly, it commutes with any $H_{t'}^2[q,p']$ or $H_{t'}^2[q',p]$ or $H_{t'}^2[q,p]$ as these terms can propagate the clock of one qubit or both qubits, but they cannot propagate the times of these clocks out of the complementary intervals $I_t$ and $I_t^c$. In other words, these last terms commute with the individual projectors $\Pi(t_q \in I_t)$,$\Pi(t_p \in I_t), \Pi(t_p \in I_t^c), \Pi(t_q \in I_t^c)$.
The commutativity implies that the eigenstates of $H$ either reside in the subspace where $H_{causal}=0$, i.e. the valid time-configuration subspace, or the subspace where $H_{causal}$ has its lowest nonzero eigenvalue which is 1. In this way we impose an energy penalty on invalid time-configurations and we can ignore them in the remainder of the analysis.

%BMT updated gap results
In the next two sections, we do the technical work of establishing both aspects of the map where the final results are expressed in Eq.~(\ref{eq:low}) and Eq.~(\ref{eq:high}). Note that the difference between $a$ and $b$ scales as $\frac{1}{DS^2}$ where $S$ is the size of the verification circuit and $D$ is its depth, if $\epsilon$ is sufficiently small. This proof is very analogous to the standard proof, first given in \cite{KSV:computation}, with similar results, but the notation and some of details are a bit more cumbersome.

\subsubsection{Yes-instance $\Rightarrow$ (almost) zero energy groundstate}
\label{sec:YES}
We assume that there exists an input witness state $\ket{\xi}$ such that the verification circuit $C_n$ has $q_{out}=1$ with probability $1-\epsilon$. We construct a low-energy state for the Hamiltonian $H$ in Eq.~(\ref{eq:defH}) as the history state, Eq.~(\ref{eq:history}), with $\ket{\phi_{in}}=\ket{\xi} \ket{y=00 \ldots 0}$. The terms $H_{in}, H_{prop}$ and $H_{causal}$ have zero energy with respect to this state, thus
\begin{eqnarray*}
\bra{\psi_{history}} H\ket{\psi_{history}}=\bra{\psi_{history}} H_{out} \ket{\psi_{history}} \nonumber \\ 
=\frac{1}{D {n \choose n/2}}\sum_{{\bf t}:t_{q_{out}}=D} \bra{\xi, 00 \ldots 0} V^{\dagger}({\bf t} \leftarrow {\bf 0}) \ket{0}\bra{0}_{q_{out}} V({\bf t} \leftarrow {\bf 0}) \ket{\xi, 00 \ldots 0}.
\end{eqnarray*}
Note that the valid times ${\bf t}$ with $t_{q_{out}}=D$ are times such that $V({\bf t} \leftarrow {\bf 0})$ is the product of a set of elementary gates which {\em includes all gates which are in the past causal cone of $q_{out}$}. Said differently, it includes all gates which are needed to produce the correct circuit outcome for the output qubit $q_{out}$. Hence
$\bra{\xi, 00 \ldots 0} V^{\dagger}({\bf t} \leftarrow {\bf 0}) \ket{0}\bra{0}_{q_{out}} V({\bf t} \leftarrow {\bf 0}) \ket{\xi, 00 \ldots 0}\leq \epsilon$. The number of ${\bf t}$ for which $t_{q_{out}}=D$ is simply ${n-1 \choose \frac{n}{2}-1}$ as fixing the time for one qubit fixes the counter $\tau$ and the first bit of the bit string $x$.
Thus 
\begin{equation}
\bra{\psi_{history}} H \ket{\psi_{history}} \leq \frac{\epsilon}{2D} \equiv a.
\label{eq:low}
\end{equation}

\subsubsection{No-instance $\Rightarrow$ ground-state energy of Hamiltonian bounded away from zero}
\label{sec:NO}
We start from the assumption that for all inputs $\ket{\xi} \ket{00 \ldots 0}_{S_{in}}$ to the verification circuit $C_n$,  
$\prob{[q_{out}=1]} \leq \epsilon$. Due to the presence of $H_{causal}$ and the fact that $H_{circuit}$ preserves the subspace of valid time-configurations, the eigenstates of $H$ in the space of invalid time configurations have energy penalty at least 1. We thus consider the spectrum of $H_{circuit}+H_{in}+H_{out}$ in the space of valid time configurations. 

We apply Lemma \ref{lem:kitaev} with $A=H_{circuit}(\{U\})$ and $B=H_{in}+H_{out}$ which have no common null-space as the quantum circuit never outputs $q_{out}=1$ for some correctly initialized input state by assumption. The final result is the following lowerbound
\begin{lem}
For a no-instance the smallest eigenvalue of the Hamiltonian $H$ can be lowerbounded as  
\begin{equation}
\lambda_1(H) \geq \Omega\left(\frac{1}{D^2 n^2}\right) \left(\frac{1}{4D}-O\left(\frac{\epsilon}{D}\right)\right) \equiv b
\label{eq:high}
\end{equation}
\label{lem:high}
\end{lem}

{\em Proof}: Theorem \ref{thm:lb} provides the lower-bound on $\lambda_1(H_{circuit})$. Consider $B$ and note that the set $\{{\bf t}\colon t_{q_{out}}=D\}$ is disjoint from the sets 
$\{{\bf t}\colon t_{p\in S_{in}}=0\}$ as we have assumed that the qubits in $S_{in}$ are in the past causal cone of $q_{out}$ thus their clocks cannot read $t=0$ while the clock of the output qubit reads $D$! This means that $\lambda_1(B) \geq 1$. To apply Lemma \ref{lem:kitaev}, we need to bound the angle between the null-spaces of $A$ and $B$.  The nullspace of $A$ only contains the history states $\psi_{history}$ in Eq.~(\ref{eq:history}).  The goal is to upperbound $\cos^2(\theta)=\max_{\psi_{history}} \bra{\psi_{history}}\Pi_B \ket{\psi_{history}}$ where $\Pi_B$ is the projector onto the nullspace of $B$.
We can write $\ket{\psi_{history}}=\alpha_I \ket{\psi_I}+\alpha_{NI} \ket{\psi_{NI}}$ where $\psi_I$ is a state which is properly initialized, i.e. $\ket{\phi_{in}^I}=\ket{\xi,00 \ldots 0}$ and $\psi_{NI}$ is some state which is not properly initialized.  We have 
\begin{eqnarray}
\bra{\psi_{history}} \Pi_B \ket{\psi_{history}}=&|\alpha_I|^2 \bra{\psi_I} \Pi_B \ket{\psi_I}+|\alpha_{NI}|^2 \bra{\psi_{NI}} \Pi_B \ket{\psi_{NI}} \nonumber \\ &+2 Re(\alpha_I \alpha_{NI}^* \bra{\psi_{NI}} \Pi_B \ket{\psi_{I}}). 
\label{eq:alltogeth}
\end{eqnarray}
We will separately determine the maximum values of $\bra{\psi_I} \Pi_B \ket{\psi_I}$ and $\bra{\psi_{NI}} \Pi_B \ket{\psi_{NI}}$ and the crossterm $|\bra{\psi_{NI}} \Pi_B \ket{\psi_{I}}|$. We start with some basic observations. The nullspace of B is a direct sum of spaces $\ker(B)=\ker(B)_{out}\oplus \ker(B)_{in}\oplus \ker(B)_{int}$ with the three orthogonal null-spaces:
\begin{eqnarray*}
\ker(B)_{out}=&{\rm span}\left(\ket{1}_{q_{out}} \ket{v} \otimes \ket{{\bf t}\colon t_{q_{out}}=D}, \forall \ket{v}\in ({\cal C}^2)^{\otimes n-1}\right) \nonumber \\
\ker(B)_{in}=&{\rm span}\Big(\ket{w} \ket{00 \ldots 0}_{S(x)} \otimes \ket{{\bf t}\colon  \forall p \in S(x),\; (t_p=0)},\nonumber \\
& \forall S(x) \neq \emptyset \subseteq S_{in}, \forall \ket{w}\in ({\cal C}^2)^{\otimes n-1}\Big) \nonumber \\
\ker(B)_{int}=&{\rm span}\left(\ket{\xi}\otimes \ket{{\bf t}\colon (\forall p, t_{p} \neq 0) \wedge (t_{q_{out}} \neq D}), \forall \ket{\xi}\in ({\cal C}^2)^{\otimes n}\right). 
\end{eqnarray*}
We have $\Pi_B=\Pi_{in}+\Pi_{out}+\Pi_{int}$ where $\Pi_{in}, \Pi_{out}$ and $\Pi_{int}$ are the projectors onto these three null-spaces. As $\Pi_{int}$ is diagonal in the ${\bf t}$-basis, we have
\begin{eqnarray*}
\bra{\psi_{history}} \Pi_{int} \ket{\psi_{history}}=\frac{|\{{\bf t}\colon (t_{q_{out}} \neq D) \wedge (\forall p \in S_{in}, t_p \neq 0)\}|}{D {n \choose n/2}},
\end{eqnarray*}
independent of initialization or the witness state. 

By assumption on the verification circuit we have for all proofs $\ket{\phi_{in}^I}=\ket{\xi,00 \ldots 0}$
\begin{eqnarray*}
\bra{\psi_{I}} \Pi_{out} \ket{\psi_I} &=
\frac{1}{D {n \choose n/2}}\sum_{{\bf t}:t_{q_{out}}=D}
 \bra{\phi_{in}^I}  V^{\dagger}({\bf t} \leftarrow {\bf 0}) \ket{1}\bra{1}_{q_{out}} V({\bf t} \leftarrow {\bf 0}\ket{\phi_{in}^I} \\
 &\leq \frac{\epsilon}{2D},
\end{eqnarray*}
where we used that {\em all} $V({\bf t} \leftarrow {\bf 0})$ with $t_{q_{out}}=D$ are evolutions which lead to the correct output of the verification circuit. This implies that for all proofs $\psi_I$, we have
\begin{equation}
\bra{\psi_I} \Pi_B \ket{\psi_I}=1-\frac{1-\epsilon}{2D}.
\label{eq:I}
\end{equation}

Consider next $\bra{\psi_{NI}} \Pi_B \ket{\psi_{NI}}$. We have $\bra{\psi_{NI}} \Pi_B \ket{\psi_{NI}} \leq \max_{\psi_{NI}} \bra{\psi_{NI}} \Pi_{out}\ket{\psi_{NI}}+ \max_{\psi_{NI}} \bra{\psi_{NI}} \Pi_{int}+\Pi_{in} \ket{\psi_{NI}}$. The first term is maximized when we assume that all improperly initialized states lead to $q_{out}=1$.
We focus on upperbounding the last term $\bra{\psi_{NI}} \Pi_{in} \ket{\psi_{NI}}$. We write
\begin{equation}
\Pi_{in}=\sum_{S \neq \emptyset \in S_{in}} \ket{00 \ldots }\bra{00 \ldots}_{S} \otimes P_{S},
\label{eq:projint}
\end{equation}
with $P_{S}$ the projector onto all $\ket{\bf t}$ for which $(\forall p\in S, t_p=0) \wedge (\forall p \in S_{in}\backslash S, t_p \neq 0)$.  
Let the state $\psi_{NI}$ be initialized to some $\ket{\phi_{in}^{NI}}=\sum_{y \neq 00 \ldots 0 \in \{0,1\}^m} \ket{\xi_y}\otimes \ket{y}_{S_{in}}$. We note that the projector $\Pi_{in}$ in Eq.~(\ref{eq:projint}) acts diagonally on the basis $\ket{y}_{S_{in}}$ which implies that the input state $\phi_{in}^{NI}$ initialized with a $\ket{y}_{S_{in}}$ which `incurs a minimal penalty' is the one which for which $\bra{\psi_{NI}} \Pi_{in} \ket{\psi_{NI}}$ is maximized. For this particular $y$, all qubits in $S_{in}$ are set to 0, except for one qubit, call it qubit $q_1$, whose state is set to 1. 
Let this particular subset of qubits which is initialized to $0$ be $T \subseteq S_{in}$ \footnote{In order to not have any dependence on the particular choice for qubit $1$, we assume for simplicity that the number of qubits in $S_{in}$ is even, that the qubits are adjacent to each other and that they all interact among each other at the first time-step.}. 
Taking $\ket{\psi_{NI}}$ initialized with $\ket{\phi_{in}^{NI}}=\ket{\xi} \ket{100 \ldots 0}_{S_{in}}$, one has: 
\begin{eqnarray*}
\bra{\psi_{NI}} \Pi_{in} \ket{\psi_{NI}}&=\sum_{\emptyset \neq S \subseteq S_{in}} \frac{{\rm Rank}(P_{S})}{D {n \choose n/2}} {\rm Tr}(\ket{10 \ldots 0}\bra{10 \ldots 0}_{S_{in}} \ket{0 \ldots 0}\bra{00 \ldots 0}_{S}) \nonumber \\
&=\sum_{\emptyset \neq S \subseteq T} \frac{{\rm Rank}(P_{S})}{D {n \choose n/2}}=\sum_{\emptyset \neq S \subseteq S_{in}} \frac{{\rm Rank}(P_{S})}{D {n \choose n/2}}-\sum_{\emptyset \neq S \in S_{in}:q_1  \in S} \frac{{\rm Rank}(P_{S})}{D {n \choose n/2}}.
\end{eqnarray*}
Note that for a properly initialized state we have
\begin{equation*}
\bra{\psi_{I}} \Pi_{in} \ket{\psi_I}=\sum_{\emptyset \neq S \subseteq S_{in}} \frac{{\rm Rank}(P_{S})}{D {n \choose n/2}}
\end{equation*}
Furthermore
\begin{eqnarray*}
\sum_{\emptyset \neq S \subseteq S_{in}:q_1 \in S} {\rm Rank}(P_{S})&=\sum_{q_1 \in S \in S_{in}} |\{{\bf t}\colon (\forall p\in S, t_p=0) \wedge (\forall p \in S_{in}\backslash S, t_p \neq 0)\}|\\
&=|\{{\bf t} \colon t_{q_1}=0\}|={n-1 \choose \frac{n}{2}-1}.
\end{eqnarray*}
This gives 
\begin{equation}
\max_{\psi_{NI}} \bra{\psi_{NI}} \Pi_B \ket{\psi_{NI}}=1-\frac{1}{2D}.
\label{eq:NI}
\end{equation}

Lastly, we bound the `crossterm' $|\bra{\psi_{NI}} \Pi_B \ket{\psi_{I}}|$. Following the slightly different proof technique in \cite{AGIK:1d}, we can write $\Pi_B=\Pi_{final}\Pi_{init}$ where $\Pi_{init}$ is the projector onto the entire nullspace of $H_{in}$ and $\Pi_{final}$ is the projector onto the null-space of $H_{out}$. The projectors $\Pi_{init}$ and $\Pi_{final}$ commute as the set $\{{\bf t}\colon t_{q_{out}}=D\}$ is disjoint from the sets $\{ {\bf t}\colon t_{p \in S_{in}}=0\}$. We have 
\begin{equation*}
|\bra{\psi_{NI}} \Pi_{final} \Pi_{init} \ket{\psi_{I}}| \leq |\bra{\psi_{NI}} \Pi_{final} \ket{\psi_I}|.
\end{equation*}
As $\Pi_{final}$ is diagonal in the basis ${\bf t}$ and a properly initialized state $V({\bf t} \leftarrow {\bf 0}) \ket{\psi_{in}^I} \otimes \ket{\bf t}$ is orthogonal to $V({\bf t} \leftarrow {\bf 0}) \ket{\psi_{in}^{NI}} \otimes \ket{\bf t}$, we can bound
\begin{eqnarray}
|\bra{\psi_{NI}} \Pi_{final} \ket{\psi_I}|\leq \nonumber \\ \frac{1}{D {n \choose n/2}}\sum_{{\bf t}:t_{q_{out}=D}} |\bra{\psi_{in}^{NI}} V^{\dagger}({\bf t} \leftarrow {\bf 0}) \ket{1}\bra{1}_{q_{out}} V({\bf t} \leftarrow {\bf 0})\ket{\psi_{in}^{I}}| \leq \frac{\sqrt{\epsilon}}{2D} 
\label{eq:cross}
\end{eqnarray}

All contributions, Eqs.~(\ref{eq:I}),(\ref{eq:NI}),(\ref{eq:cross}) together with Eq.~(\ref{eq:alltogeth}) give
\begin{equation}
\bra{\psi_{history}} \Pi_B \ket{\psi_{history}} \leq 1-\frac{1}{2D}+\frac{\epsilon}{2D} +\frac{\sqrt{\epsilon}}{D},
\end{equation}
which is bounded away from $1$  by approximately $\frac{1}{2D}$ for exponentially small (in $n$ or $D$) $\epsilon$.
Using Lemma \ref{lem:kitaev} then gives Eq.~(\ref{eq:high}). $\Box$.

\subsection{Clock Realizations}
\label{sec:clock}

The space-time circuit Hamiltonians $H_{circuit}$ used so far are not $O(1)$-local Hamiltonians, --they are not sums of terms each of which acts on O(1) qubits non-trivially,-- as the clock of each qubit is realized by a $O(\log D)$-qubit register. In order to prove that the lowest eigenvalue problem for $O(1)$-local Hamiltonians is QMA-complete, one can realize such clock as a pulse or domain wall clock (see e.g. \cite{nagaj:circular}). In particular for the domain-wall clock introduced by Kitaev \cite{KSV:computation}, terms such as $\ket{t}\bra{t-1}$ are 3-local. For the QMA-application, one then considers a Hamiltonian $H=H_{circuit}+H_{in}+H_{out}+H_{causal}+H_{clock}$ where $H_{clock}$ gives a $O(1)$ penalty to any state of the time-registers which does not represent time. This implies that the lowest-energy states are in the space where the time-registers do represent time and one applies the arguments in the previous sections to this subspace. Using the domain wall clock in the space-time circuit-to-Hamiltonian construction gives rise to $8$-local terms as $\ket{t,t}\bra{t-1,t-1}$ is $6$-local. Similarly, the term $H_{causal}$ translates into a 4-local term as a term of the form $\ket{t}\bra{t}$ is 2-local for a domain wall clock, e.g. \cite{nagaj:circular}. This implies that this use of the space-time circuit-to-Hamiltonian construction is less efficient in terms of locality than the Feynman-Kitaev construction which is $5$-local.

\subsection{QMA-completeness of two-dimensional interacting fermions}
\label{sec:QMA2Dfermions}

\begin{figure}[htb]
\centering
\includegraphics[width=1\hsize]{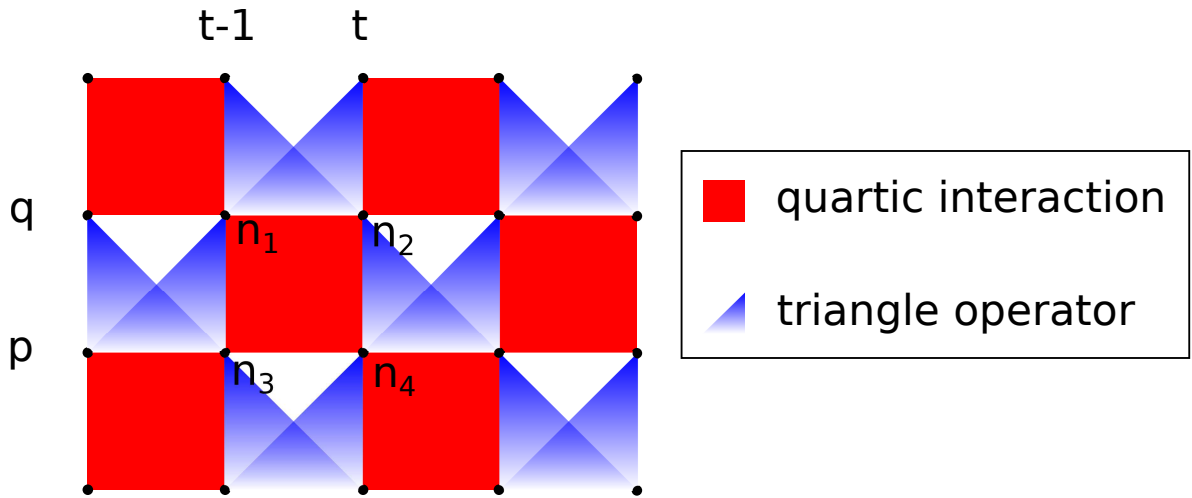} 
\caption{The black dots are fermionic sites, each with two modes (an $\uparrow$ or $\downarrow$ state, say). The (red) squares represent the quartic gate interactions and the (blue) triangle operators penalize invalid fermionic configurations (invalid time-configurations). A (blue) triangle operator with top corner $a$ and bottom corners $b$ and $c$ equals $n_a(1-n_b-n_c)$. The lattice has periodic boundary conditions in both directions.}
\label{fig5}
\end{figure}

We can also prove QMA-completeness for the fermionic model of \cite{MLM:qadiabatic} (\cite{thesis:breuckmann}) which indirectly realizes a pulse clock for each qubit $q$. The terms of the circuit Hamiltonian are in Eq.~(\ref{eq:CU}) in Section \ref{sec:relateMLM}. Note that we can only represent two-qubit gates which are controlled-$U$ operations. However, given a supply of qubits initialized to the state $\ket{1}$, a one-dimensional quantum circuit with only such controlled-$U$ gates is universal. The circuit Hamiltonian will correspond to that of an interacting fermion model in two spatial dimensions with periodic boundary conditions in both directions (a torus), as we work with the circular time circuit-to-Hamiltonian construction. Aside from the circuit Hamiltonian one needs the fermionic equivalent of the terms $H_{in}$, $H_{out}$ and $H_{causal}$. To represent the input state $\ket{00\ldots 0}_{S_{in}}$, one takes
\begin{equation*}
H_{in}=\sum_{q \in S_{in}} b_0^{\dagger}[q] b_0[q],
\end{equation*}
such that the modes $b_0[q]$ (corresponding to those qubits being in the state $\ket{1}$ at time 0) are never occupied. If we translate this back to qubits, this corresponds to the term $H_{in}$ in Eq.~(\ref{eq:hin}). Similarly, for $H_{out}$, Eq.~(\ref{eq:hin}), one takes 
\begin{equation*}
H_{out}=a_D^{\dagger}[q_{out}]a_D[q_{out}]. \nonumber
\end{equation*}
Lastly, $H_{causal}$ (given in \cite{MLM:qadiabatic}) is the fermionic equivalent of Eq.~(\ref{eq:caus}). For a gate in the original quantum circuit at time $t$ between qubits $q$ and $p$, one can take
\begin{equation}
H_{causal}([q,p],t)=n(t_q \in I_t) n(t_p \in I_t^c)+n(t_p \in I_t) n(t_q \in I_t^c),
\label{eq:causmlm}
\end{equation}
where $n(t_q \in I_t)=\sum_{t_q \in I_t} n_{t_q}[q]$ with number operator $n_{t_q}$ (previously defined in Sec.~\ref{sec:relateMLM}). Again $H_{causal}$ commutes with all other terms $H_{in},H_{out}$ and $H_{circuit}$.
This form of $H_{causal}$ is not local on the two-dimensional lattice however. If we wish to prove QMA-completeness of the ground-state energy problem of a two-dimensional interacting fermion model, then one can replace $H_{causal}$ by a local version $H_{causal}^{loc}$. The idea is that the valid time-configurations of the quantum circuit in Fig.~(\ref{fig2})(b) are very constrained.
%NB removed ", or the space-time is tightly knit"
Consider Fig.~(\ref{fig5}). In between all two-qubit gates, --which themselves form a checkerboard pattern--, one places two triangle operator constraints. The triangle operator between three fermionic sites $a$, $b$ and $c$ with control site at the top labeled $a$, see Fig.~(\ref{fig5}) reads $H_{triangle}=n_a(1-n_b-n_c)$. It is important to note that we work in the Fock space where $N[q]=1$ which means that $\langle n_b+n_c \rangle \leq 1$ and $H_{triangle} \geq 0$ for the triangle operators in the picture. The zero energy subspace of $H_{triangle}$ is the direct sum of the Fock-space with $n_a=0$, the space with $n_a=1$ and $n_b=1$, and the space with $n_a=1$ and $n_c=1$. Thus the triangle operator expresses the constraint that {\em if there is a particle at $a$, there should also be a particle at $b$ or $c$}. In the spaces between the gates, one puts two triangle operators. Note that the triangle operators all commute as all number operators $n_t[q]$ mutually commute. 

It is not hard to see that all triangle operators have energy zero if and only if the fermionic Fock states represent a valid time-configuration. In addition, we want to establish that the sum over all triangle operators commutes with $H_{circuit}$, $H_{in}$ and $H_{out}$. When this is the case, the lowest invalid Fock state has at least energy $1$ and thus in order to determine the lowest nonzero eigenvalue of $H$, one only needs to look at the space of valid Fock states. Consider a gate term $H_t^{CU}[q,p]$ with qubits $q,p$ as control and target qubits in Eqs.~(\ref{eq:CNOT}),(\ref{eq:CU}), as in Fig.~(\ref{fig5}) with the number operators $n_1$, $n_2$ and $n_3$ and $n_4$ at the corners of the gate. We wish to show that all triangle operators commute with $H_t^{CU}[q,p]$. We consider the gate interaction $H_t^{CU}[q,p]$ on the states partially labeled by $n_1,n_2,n_3, n_4,\{n_{else}\}$ where $\{n_{else}\}$ are the number operators for all the other fermionic sites on the lattice (the full state specification includes the spin-degree but is not relevant for the next arguments).

Due to the $\forall q,\;N[q]=1$ constraint, some of these $n_{else}$ are constrained depending on $n_1,\ldots, n_4$: in particular we only have $(n_1,n_2,n_3,n_4)=(1,0,1,0),(1,0,0,1),(0,1,0,1),(0,1,1,0),(0,0,1,0),(0,0,0,1),(1,0,0,0),(0,1,0,0)$ and $(0,0,0,0)$. $H_t^{CU}[q,p]$ has nontrivial action only in the subspace where $(n_1,n_2,n_3,n_4)=(1,0,1,0)$ and $(n_1,n_2,n_3,n_4)=(0,1,0,1)$, for all other $(n_1,n_2,n_3,n_4)$  states it has zero energy. This means that the operators $n_1+n_2$, $n_3+n_4$ and $n_1 n_3+n_1 n_4$ commute with the gate interaction. The four triangle operators above and below the gate, see Fig.~(\ref{fig5}) involves only symmetric combination such as $n_1+n_2$ and $n_3+n_4$  and thus commute. The sum of the two triangle operators left and right to the gate can be written as $(n_1+n_2)-(n_1 n_3+n_2 n_4)-n_1 n_5 -n_2 n_6$ where the first two terms in $()$ are conserved quantities and thus commute. The last two terms commute separately as they only have support on the null-space of the gate interaction. Similarly the triangle operator, either on the left or the right of the gate, commutes with the gate interaction as the only term which involves, say $n_3$, is supported on the null-space of the gate interaction. Note that the triangle operators also commute with $H_{in}$ and $H_{out}$.
This means that the fermionic Hamiltonian $H=H_{circuit}+H_{in}+H_{out}+H_{causal}^{loc}$ is a quartic fermion Hamiltonian involving spin-1/2 fermionic sites. The quartic interaction involves at most 4 fermionic sites on a square lattice, see Fig.~(\ref{fig5}).

%NB removed comment on AQC
The mappping from a 2D fermionic Hamiltonian onto the space-time circuit Hamiltonian $H_{circuit}$ assumes that there is at most one fermion per qubit $q$, i.e $N[q]=1$, see the mapping in Section \ref{sec:relateMLM}. This means that the arguments above and in the last sections show that the problem of deciding whether there is a state with energy less than or equal to $a$ or larger than or equal to $b$ ($|a-b| \geq \frac{1}{{\rm poly}(n)}$) for a two-dimensional interacting fermion Hamiltonians $H$ on a torus, {\em in the sector where $\forall q, N[q]=1, N[q]=\sum_{t \in Z_{2D}} n_t[q]$} is QMA-complete. This result goes beyond the perturbative approach used in \cite{sv:qma} as all terms in the Hamiltonians here are of strength $O(1)$. Considering eigenvalues of fermionic problems restricted to sectors with fixed number of fermions is not unnatural as fermion number is a conserved quantity in physical systems and one can tune a physical system such as a quantum dot so that one excess electron (above the Fermi energy) is available for interactions.
Alternatively, we add a nonlocal penalty term $H_{clock}$ to the Hamiltonian which enforces $N[q]=1$, e.g. $H_{clock}=\sum_q (N[q]-1)^2$. However, as has been observed before \cite{nagaj:circular}, it is not clear how to enforce this constraint in a local one-dimensional manner (without making the vacuum state without fermions always have the lowest energy).

We note that these results also can be stated in terms of only qubits instead of fermions (using the Jordan-Wigner transformation). The terms $H_{in}, H_{out},H_{caus}^{loc}$ remain local terms under this transformation. However the pulse clock condition $\forall q, N[q]=1$ is somewhat less natural. 

\subsection{Quantum adiabatic computation}
\label{sec:fermrevis}

We consider how the results in this paper can be used for simulating a quantum circuit by a quantum adiabatic computation. 
 One assumes that the quantum circuit which we wish to simulate by an adiabatic computation is efficient, i.e. $L={\rm poly}(n)$ where ${\rm poly}(n)$ is some polynomial in $n$. A simple way to go from the circuit Hamiltonian to an adiabatic algorithm is to construct a continuous family of circuit Hamiltonians $H_{circuit}(U_1(\epsilon), \ldots, U_L(\epsilon))=H_{circuit}[\epsilon]$ depending on a parameter $\epsilon \in [0, 1]$. For $\epsilon=0$, we have $\forall i\; U_i(\epsilon=0)=I$ while for $\epsilon=1$, we have $U_i(\epsilon=1)=U_i$ such that we smoothly interpolate between $I$ and $U_i$ for intermediate values of $\epsilon$ \cite{BT:stoq} (Such smooth deformations always exists as one can continuously deform any element to $I$ in a Lie-group $U(n)$) \footnote{In the more standard construction in \cite{ADLLKR:adia} the intermediate Hamiltonians on the adiabatic path are linearly interpolating between initial and final Hamiltonian.}. 

The adiabatic computation starts in the groundstate of $H_{circuit}[\epsilon=0]$ and $\epsilon$ is gradually increased to evolve to the groundstates of $H_{circuit}[\epsilon \neq 0]$. The smoothness in the interpolation is required such that first and second-derivatives of $H_{circuit}[\epsilon]$ with respect to $\epsilon$ are polynomially bounded in $n$, so that the explicit formulation of the quantum adiabatic theorem in e.g. \cite{AR:adia} applies. In order to use the space-time Hamiltonian construction for quantum adiabatic computation one has to (i) bound the gap above the ground-state for the quantum adiabatic path $H_{circuit}[\epsilon]$, $\epsilon \in [0,1]$. Since $H_{circuit}[\epsilon]$ is unitarily related to $H_{circuit}[\epsilon=0]$, one just needs to bound the gap of $H_{circuit}[\epsilon=0]$. Secondly, one has to show that one can prepare the ground-state of the initial Hamiltonian $H_{circuit}[0]$ efficiently and thirdly show that one can read out the output state of the quantum circuit from the ground-state of the final Hamiltonian $H_{circuit}[1]$ on the adiabatic path. 

Theorem \ref{thm:lb} shows that the gap of the circuit Hamiltonian for efficient one-dimensional quantum circuits is lowerbounded appropriately, by some $\frac{1}{{\rm poly}(n)}$. Together with the unitary relation between the fermionic model and the qubit circuit Hamiltonian, this shows that the two-dimensional interacting fermionic (or qubit) model in Section \ref{sec:relateMLM} could be used for quantum adiabatic computation, as proposed in \cite{MLM:qadiabatic}. However,  one still has to show how one can prepare the initial history state (with $U=I$) as output state from another adiabatic path, as in \cite{ADLLKR:adia}, and prove that this adiabatic path has a $1/{\rm poly}(n)$ gap everywhere. In \cite{MLM:qadiabatic} the authors propose to execute the quantum adiabatic computation by gradually increasing the strength of the propagating part of each $H_t$ (by the parameter $\lambda$). However, the gap of this adiabatic path is not fully analyzed in \cite{MLM:qadiabatic, MMC:scaling} and goes beyond the results in this paper. 

If one measures the time-configuration in the history state, the total probability to measure a configuration ${\bf t}$ in which a qubit $q$ has $t_q=D$ is $\frac{1}{2D}$. This can be amplified to a constant by padding the quantum circuit with $I$ gates as in the Feynman-Kitaev construction. A different question is how one obtains the correct output for all the qubits from the history state. In \cite{TV:walk} we will give arguments why this probability scales as $\frac{1}{{\rm poly}(n)}$ when $D \gg n$.

\section{Discussion}
%NB
%It would be interesting to do a similar analysis for two-dimensional quantum circuits with periodic boundary conditions in both space directions, using the circular-time construction. For two-dimensional quantum circuits, the vertices of the underlying graph $G$ can be labeled by a 1-dimensional object (the spatial boundary of the two-dimensional quantum circuit) together with a set of internal degrees of freedom which represent a membrane with a fixed boundary.

%The reason to use the circular-time construction is that the gap analysis for the space-time circuit Hamiltonian with open boundaries in time is much more involved. The internal state-space which is now represented by $n$ qubits with $\sum_i Z_i=0$ is then different at the two time-boundaries of the circuit as strings cannot cross the boundary. Using a combination of the Markov chains techniques in \cite{MR:mixing} and results on random walks on the space of Dyck paths \cite{bravyi+:dyck}, it may be possible to bound the gap of the circuit Hamiltonian.

We note that the circuit Hamiltonian in the altered representation, Eq.~(\ref{eq:sameH}), could be directly used as a realization of a one-dimensional translationally-invariant cellular automaton circuit. For such a cellular automaton circuit, we assume that the same set of two-qubit gates is applied at every depth. This would imply that the circuit Hamiltonian is that of a purely {\em one-dimensional} system where one of the local degrees of freedom is of dimension $D$ \footnote{However, one cannot work with circular time {\em and} keep the circuit completely translationally invariant as the circular time construction then requires one to add a single I-layer of gates which is not feasible under the cellular automaton assumption.}.

Another applicaton of our analysis is a different proposal for the implementation of universal quantum computation using a time-independent two-dimensional interacting fermion system. In \cite{nagaj:circular} the standard Feynman-Kitaev construction and its spectral analysis were directly used to show how to run a quantum computation using a time-indendepent Hamiltonian. Here one expects that by initializing the fermions around the $t=0$ modes and letting them evolve for a random time within a certain window whose length scales polynomially with $n$ and $D$ one can, with high probability, measure the output state of 1 qubit of the original one-dimensional quantum circuit. 

\section{Acknowledgements} We would like to thank David DiVincenzo and Norbert Schuch for useful discussions concerning the model. BMT would like to thank Guillaume Aubrun for pointing out Ref.~\cite{BS:randomwalk_on_randomwalk}. 
BMT thanks the Isaac Newton Institute for Mathematical Sciences in Cambridge, UK for hosting the program on Mathematical Problems in Quantum Information Theory where some of this work was completed. BMT is happy to acknowledge funding through the European Union via QALGO FET-Proactive Project No. 600700.

\section*{References}
\bibliographystyle{unsrt}
%\bibliography{refsQMA}

\end{document}